\documentclass{article}

\usepackage{arxiv}

\usepackage[utf8]{inputenc} 
\usepackage[T1]{fontenc}    
\usepackage{hyperref}       
\usepackage{url}            
\usepackage{booktabs}       
\usepackage{amsfonts}       
\usepackage{nicefrac}       
\usepackage{microtype}      
\usepackage{lipsum}

\usepackage{amsmath}
\usepackage{caption}
\usepackage{subcaption}
\usepackage{xcolor}
\usepackage{graphicx}
\usepackage[superscript]{cite}
\usepackage{sidecap}
\usepackage{multirow}

\title{Solubility prediction of organic molecules\\with molecular dynamics simulations}

\author{
  Zoran ~Bjelobrk\\
  Institute of Energy and Process Engineering\\
  ETH Z\"urich, CH-8092, Switzerland\\
  \And
  Dan ~Mendels\\
  Pritzker School of Molecular Engineering\\
  University of Chicago\\
  Chicago, Illinois 60637, United States\\
  \And
 Tarak Karmakar\\
  Istituto Italiano di Tecnologia (IIT)\\
  Via Morego, 30, 16163 Genova GE, Italy\\
  \And
  Michele Parrinello\thanks{michele.parrinello@iit.it}\\
  Istituto Italiano di Tecnologia (IIT)\\
  Via Morego, 30, 16163 Genova GE, Italy\\
  \And
  Marco Mazzotti\thanks{marco.mazzotti@ipe.mavt.ethz.ch}\\
  Institute of Energy and Process Engineering\\
  ETH Z\"urich, CH-8092, Switzerland\\
}

\begin{document}
\maketitle

\begin{abstract}
We present a molecular dynamics simulation method for the computation of the solubility of organic crystals in solution.
The solubility is calculated based on the equilibrium free energy difference between the solvated solute and its crystallized state at the crystal surface kink site.
In order to efficiently sample the growth and dissolution process, we have carried out well-tempered Metadynamics simulations with a collective variable that captures the slow degrees of freedom, namely the solute diffusion to and adsorption at the kink site together with the desolvation of the kink site.
Simulations were performed at different solution concentrations 
using constant chemical potential molecular dynamics
and the solubility was identified at the concentration at which the free energy values between the grown and dissolved kink states were equal.
The effectiveness of this method is demonstrated by its success in reproducing the experimental trends of solubility of urea and naphthalene in a variety of solvents.
\end{abstract}


\section{Introduction}

Molecular solubility is a crucial indicator of functionality in pharmaceutical drugs. Not only does it dictate the driving force for crystallization of drugs in their purification process, but it also dictates their bio-availability in the body upon their intake. To date, determining the solubility of candidate drug molecules has been primarily done using experiments. However, experiments often are time consuming and for some systems expensive and tedious.
Given these deficiencies molecular dynamics (MD) simulations have been deemed to be a potentially useful tool for predicting the solubilities of candidate drug molecules and hence providing guidance in screening processes.

In approaches involving MD simulations, there are two major ways to extract the solubility for a two-phase crystal-liquid system. Either indirectly through thermodynamic cycles, where the solubility is obtained by the combination of individual simulations of the crystallized and fluid phases \cite{Ferrario2002, Paluch2013, Mester2015, Benavides2016, Li2017, Khanna2020}, or through direct coexistence simulations, where the crystal surface is exposed to the solution to sample growth and dissolution events \cite{Manzanilla-Granados2015, Kolafa2016, Espinosa2016, Asadi2015}. Although direct coexistence simulations are conceptually a more straight forward approach compared to thermodynamic cycles, they suffer from a time scale limitation. Namely, the time scale required to obtain sufficient statistics for the estimation of the solubility of a given system is considerably longer than the time scale which can be simulated with present day computational capabilities \cite{Manzanilla-Granados2015, Kolafa2016, Espinosa2016}.
To overcome this limitation in the context of crystallization, enhanced sampling was introduced with considerable success to study layered growth and homogeneous nucleation of small organic molecules in solution \cite{Salvalaglio2013, Giberti2015, Salvalaglio2015b}. Here, we shall follow this approach with the aim of calculating the solubilities for given solute-solvent systems. 

\begin{figure}[!htbp]
\begin{center}
\includegraphics[width=15cm]{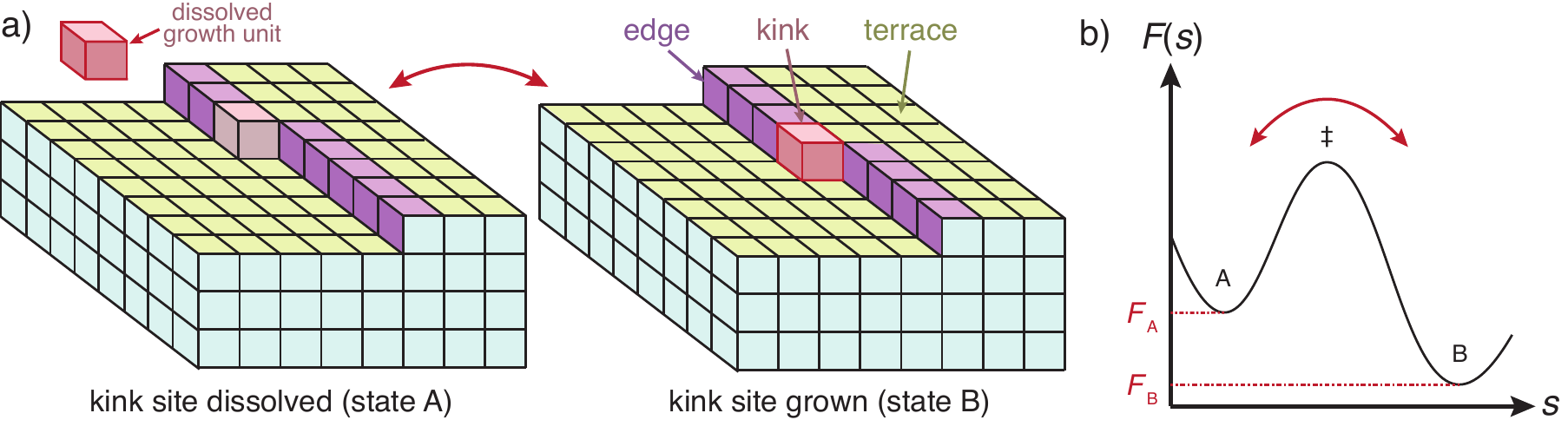}
\end{center}
\caption{a) Scheme of the crystal surface with an unfinished layer comprising a kink site; the growth units are shown as cubes. The cube with red contours indicates a growth unit which is either dissolved in solution (state A) or incorporated into the kink site (state B). Growth units at the surface are colored in green for terraces, violet for edges, and red for kink sites. b) Corresponding scheme of the free energy profile along the reaction coordinate, $s$, which describes appropriately the two states A and B separated by transition state $\ddagger$. The energy difference between the grown and dissolved states, $\Delta F = F_\text{B} - F_\text{A}$, defines whether the solution is undersaturated, $\Delta F > 0$, at solubility, $\Delta F = 0$, or supersaturated, $\Delta F < 0$.} \label{fig:kink_site}
\end{figure}

We focus on the growth and dissolution process of solute molecules at kink sites, which are sites that are located at the ends of unfinished molecule rows of crystal surface edges \cite{Kossel1927, Stranski1928, Burton1951}.
The study of kink site growth and dissolution allows us to extract the free energy difference between a solute molecule in the state of the crystal and its dissolved state in solution, while excluding surface free energy differences. 
It is also a natural choice, because at solution concentrations around solubility, growth and dissolution at these sites is the rate limiting crystallization step; \cite{Snyder2007, Li2016b} by modeling and simulating it we are therefore able to extract the solubility.
The scheme in Figure \ref{fig:kink_site}a) illustrates the crystal surface exposing a kink site to the solution, showing on the left the dissolved kink site (state A) and on the right the crystallized kink site (state B). The cubes indicate growth units (solute molecules), whereby the growth unit undergoing the phase transition is colored in red.
As the kink site grows by one growth unit, the free energy of the system changes by the energy difference between that corresponding to having the growth unit incorporated in the crystal lattice and that corresponding to the growth unit dissolved in solution. The surface free energy remains constant in this process, since the kink growth regenerates a new kink site and therefore preserves the number of crystalline growth units at edges, kinks or in terraces as illustrated in Figure \ref{fig:kink_site}a).

The corresponding scheme of the free energy surface, $F(s)$, of the kink growth process is shown in Figure \ref{fig:kink_site}b), where $s$ corresponds to a reaction coordinate which captures the dissolved (A) or crystalline (B) kink site states. The symbol $\ddagger$ labels the transition state, whose being a local maximum in energy indicates that growth and dissolution are activated processes. The energy difference between the two states, $\Delta F = F_\text{B} - F_\text{A}$, for a given solute mole fraction defines whether the solution is undersaturated, $\Delta F > 0$, at solubility, $\Delta F = 0$, or supersaturated, $\Delta F < 0$ \cite{Snyder2009, Li2016a, Li2016b, Tilbury2016}.

The growth and dissolution of molecules at the kink site is enhanced through well-tempered Metadynamics (WTMetaD) \cite{Barducci2008}. For this work we have developed a collective variable (CV), which captures the slow degrees of freedom for the kink site growth. The biased simulations allow us to obtain sufficient number of crossings between the states of grown and dissolved kink site to be able to resolve the difference in free energies for the different solution concentrations, which further allows us to identify the solubility with remarkable accuracy.

As the kink site grows, the solution is depleted, which is especially significant for low solution concentrations. To prevent solution depletion, the solute concentration in the region adjacent to the crystal surface is kept constant using the constant chemical potential molecular dynamics (C$\mu$MD) method developed by Perego \emph{et al.} \cite{Perego2015}.

The methodology presented here is applied to two organic species, namely urea and naphthalene. Particularly, we aim at predicting the solubility of these species in different solvents in order to capture the trends observed in experiments.

\section{Collective variables (CVs)} 

\subsection{CV for enhanced kink growth simulations}

To overcome the time scale limitation of kink site growth, we use enhanced sampling with the WTMetaD method. An integral part of a CV-based enhanced sampling method like WTMetaD is the appropriate choice of CVs; they are functions of atomic coordinates, and embed the system's slow degrees of freedom.\cite{Torrie1977, Laio2002}
In WTMetaD, a time dependent bias potential is constructed as a function of these CVs, with the aim to discourage frequently visited states and encourage the system to overcome free energy barriers.

For a kink site to grow, a solute molecule needs to diffuse to and adsorb at the kink site. During the adsorption event, the solute and kink site need to undergo partial desolvation. Both diffusion and adsorption are often rate-limiting \cite{Li2016a}. In the following, we systematically define a set of functions that are used to describe each of these phenomena.

\begin{figure}[!htbp]
\begin{center}
\includegraphics[width=8cm]{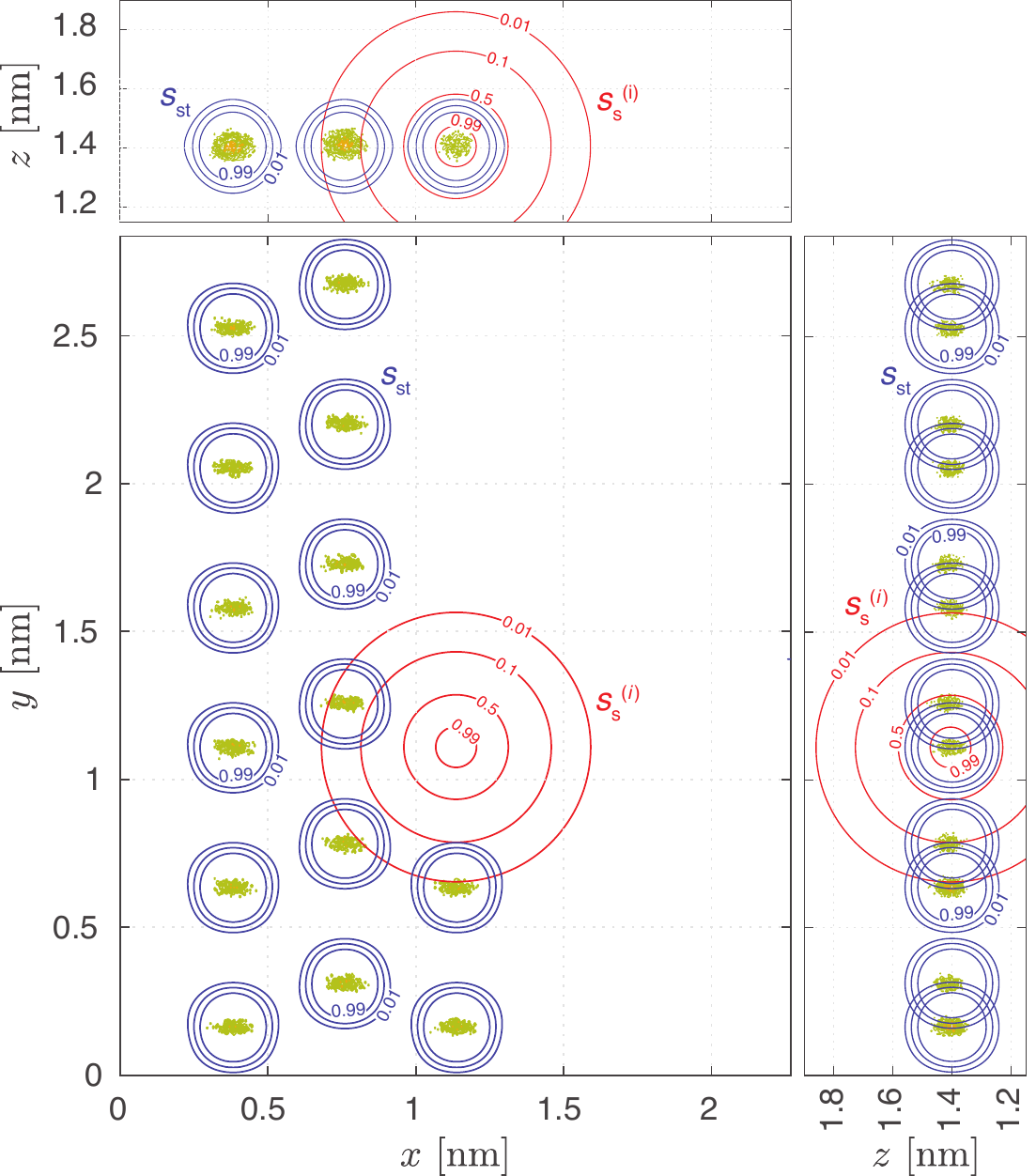}
\end{center}
\caption{Histogram of urea carbon atom positions (yellow dots) of the unfinished layer sampled from 100'000 time frames (100 ns of simulation) projected along all three spatial directions. The red contour lines correspond to the Gaussian like function, $s_\text{s}$, which is a part of the biased CV, $s_\text{b}$. $s_\text{s}$ is introduced to accelerate the solute diffusion and adsorption/desorption in the biased simulations. The blue contour lines correspond to the surface structure CV, $s_\text{st}$, through which a harmonic potential is introduced to prevent the dissolution of the unfinished layer.} \label{fig:s_sw_st}
\end{figure}

To describe the solute diffusion and its adsorption and desorption at the kink site, we define the following function:
\begin{equation}
s_\text{s} = \sum_i \exp \left( -\frac{|\mathbf{r}_i - \mathbf{r}_\text{k}|^2}{2\sigma_\text{s}^2} \right), \end{equation}
which is comprised of sums of Gaussian like bell curves summed over each solute molecule $i$ in solution or at the biased kink site.
Vectors $\mathbf{r}_i$ and $\mathbf{r}_\text{k}$ correspond to the positions of solute molecule $i$ and of kink site, respectively; $\sigma_\text{s}$ is the variance of the Gaussian function. Figure \ref{fig:s_sw_st} shows the contour lines of $s_\text{s}$ (in red) at the biased kink site of the unfinished layer of urea projected along all three spatial directions. By an appropriate choice of $\sigma_\text{s}$, the non-zero function values of $s_\text{s}$ extend into the liquid phase. This allows us, by applying bias, to push solute molecules from the liquid towards the kink site and vice versa, enhancing diffusion as well as adsorption and desorption.

To describe the solvent's desorption and adsorption at the kink site, we introduce a function with the same functional form as $s_\text{s}$:
\begin{equation}
s_\text{l} = \sum_j \exp \left( -\frac{|\mathbf{r}_j - \mathbf{r}_\text{k}|^2}{2\sigma_\text{l}^2} \right),
\end{equation}
where we sum over all solvent molecules $j$. Vector $\mathbf{r}_j$ corresponds to the position of solvent molecule $j$. The bell curve width $\sigma_\text{l}$ is chosen such that $s_\text{l}$ possesses non-zero values only at the solvent's adsorption site at the kink. Joswiak \emph{et al.} reported a similar function to $s_\text{l}$ used in their enhanced sampling simulations for water desorption and adsorption at kink sites of a rock salt \cite{Joswiak2018a, Joswiak2018b}.
In contrast to the referenced work, we employ this functional form for both the solute and the solvent. 

The CV, $s_\text{b}$, used for the WTMetaD simulations in this work is a function of $s_\text{s}$ and $s_\text{l}$:
\begin{equation}
s_\text{b} = w_\text{s} (s_\text{s}^{\chi_\text{s1}} + s_\text{s}^{\chi_\text{s2}}) + w_\text{l} s_\text{l}^{\chi_\text{l}},
\end{equation}
where $w_\text{s}$ and $w_\text{l}$ are scalar weights that are obtained by harmonic linear discriminant analysis \cite{Mendels2018a, Piccini2018, Mendels2018b, Brotzakis2019}.
To improve the WTMetaD sampling performance, $s_\text{s}$ is potentiated through positive scalar exponents $\chi_\text{s1} < 1$ and $\chi_\text{s2} > 1$. This variable transformation helps us to map the corresponding free energy surface $F: s_\text{s} \rightarrow s_\text{s}^{\chi_\text{s1}} + s_\text{s}^{\chi_\text{s2}}$ into a function with wider local minima, which otherwise would be too narrow for efficient bias deposition performance. Also $s_\text{l}$ is transformed similarly, so as the mapping of the free energy surface $F: s_\text{l} \rightarrow s_\text{l}^{\chi_\text{l}}$ leads to a wider well of the local energy minimum at $s_\text{l} = 0$. These transformations follow the approach utilized earlier. \cite{Bjelobrk2019,Rizzi2020}

It is worth noting that $s_\text{b}$ does not take into account the reorientation of the solute at the kink site, which might be a further rate limiting step for organic molecules exhibiting a more complex structure\cite{Chernov1998}.

\subsection{Surface structure CV}

While growing the kink sites that are present in a unfinished layer, it is likely that some of the crystalline molecules of the unfinished layer dissolve. This might lead to undesired difficulties in the sampling of growth and dissolution events at the kink site. 

To prevent such dissolution, a harmonic potential is introduced. This potential is defined as a function of a surface structure CV, $s_\text{st}$, whose value for each molecule is 1 if the molecule in the unfinished layer is in a lattice position or 0 otherwise (see the supporting information (SI) for further details).
In this way, $s_\text{st}$ counts the total number of molecules in the unfinished layer.
Figure \ref{fig:s_sw_st} shows the contour lines of $s_\text{st}$ projected along all three spatial directions together with the histogram of the molecule centers of the unfinished layer for the case of urea. Note that we should not alter the natural lattice vibrations of the crystalline molecules by applying the harmonic potential. This is achieved by setting the parameters of $s_\text{st}$ such that its  potential only counteracts the molecule's motion out of its lattice position. The amplitudes of the natural lattice vibrations are known from unconstrained simulations.

\subsection{Crystallinity CV}

The biased CV $s_\text{b}$ discussed in Section 2.1 does not suffice to define the kink site crystallinity since it lacks information about the solute molecule's orientational ordering. To account for this, we have introduced a pair of CVs, $s_\text{c1}$ and $s_\text{c2}$, defined using the logistic function as follows:
\begin{equation}
s_{\text{cj}} = \sum_i \left(1 - \frac{1}{1+\exp(-\sigma_\text{c}(|\mathbf{r}_{\text{cj},i}-\mathbf{\bar{r}}_\text{cj}|-d_\text{c}))}\right),\;\; (j=1,2).
\end{equation}
These CVs discriminate whether the atom positions $\mathbf{r}_{\text{c1},i}$ and $\mathbf{r}_{\text{c2},i}$ of solute molecule $i$ at the kink site are at their concordant crystal lattice positions, namely $\mathbf{\bar{r}}_\text{c1}$ and $\mathbf{\bar{r}}_\text{c2}$. Using two appropriate atom positions allows taking into account the molecule's orientation at the kink site as well, as requested. Moreover, $d_\text{c}$ is the position of the logistic function's step and is set roughly at the distance between the crystalline lattice position and the kink growth region of transition, i.e. the region where the biased kink site is neither fully dissolved nor fully crystalline. $\sigma_\text{c}$ is the steepness of the step and is chosen so as the values of the switching function, which are larger than 0 but smaller than 1, correspond roughly to the molecule's positions that are contained within the region of transition.
For $s_\text{c1}$ and $s_\text{c2}$, values around 0 correspond to a fully dissolved biased kink site while values around 1 correspond to a crystalline biased kink site (see SI for further details).

\section{Computational Details}

\subsection{Force fields}

The general AMBER force field (GAFF) \cite{Wang2004, Wang2006} with full atomistic description was used for all molecular species considered in this work. The force field parameters of naphthalene are reported in a previous paper \cite{Bjelobrk2019}. For naphthalene, the electrostatic potential was calculated using Gaussian 09 \cite{Gaussian09} at the B3LYP/6-31G(d,p) level and the atom partial charges were fitted with the restrained electrostatic potential method \cite{Bayly1993, daSilva2012}. The force field parameters of all other molecules studied in this work were taken from the literature. \cite{vanderSpoel2012}

\subsection{Simulation runs}

The simulations were performed with Gromacs 2016.5 \cite{Berendsen1995, Lindahl2001, Spoel2005, Hess2008b, Abraham2015} patched with a private version of Plumed 2.5.0 \cite{Tribello2014}. The temperature was kept constant with the velocity rescaling algorithm \cite{Bussi2009}.
The non-bonded electrostatic interactions cutoff was set to 1 nm. For the long-range electrostatics, the Ewald particle mesh algorithm \cite{Darden1993} was used. To run simulations at an integration time step of 0.002 ps, the LINCS algorithm was used to constrain the hydrogen bonds \cite{Hess2008b, Hess2008a}.

\begin{figure}[!htbp]
\begin{center}
\includegraphics[width=7cm]{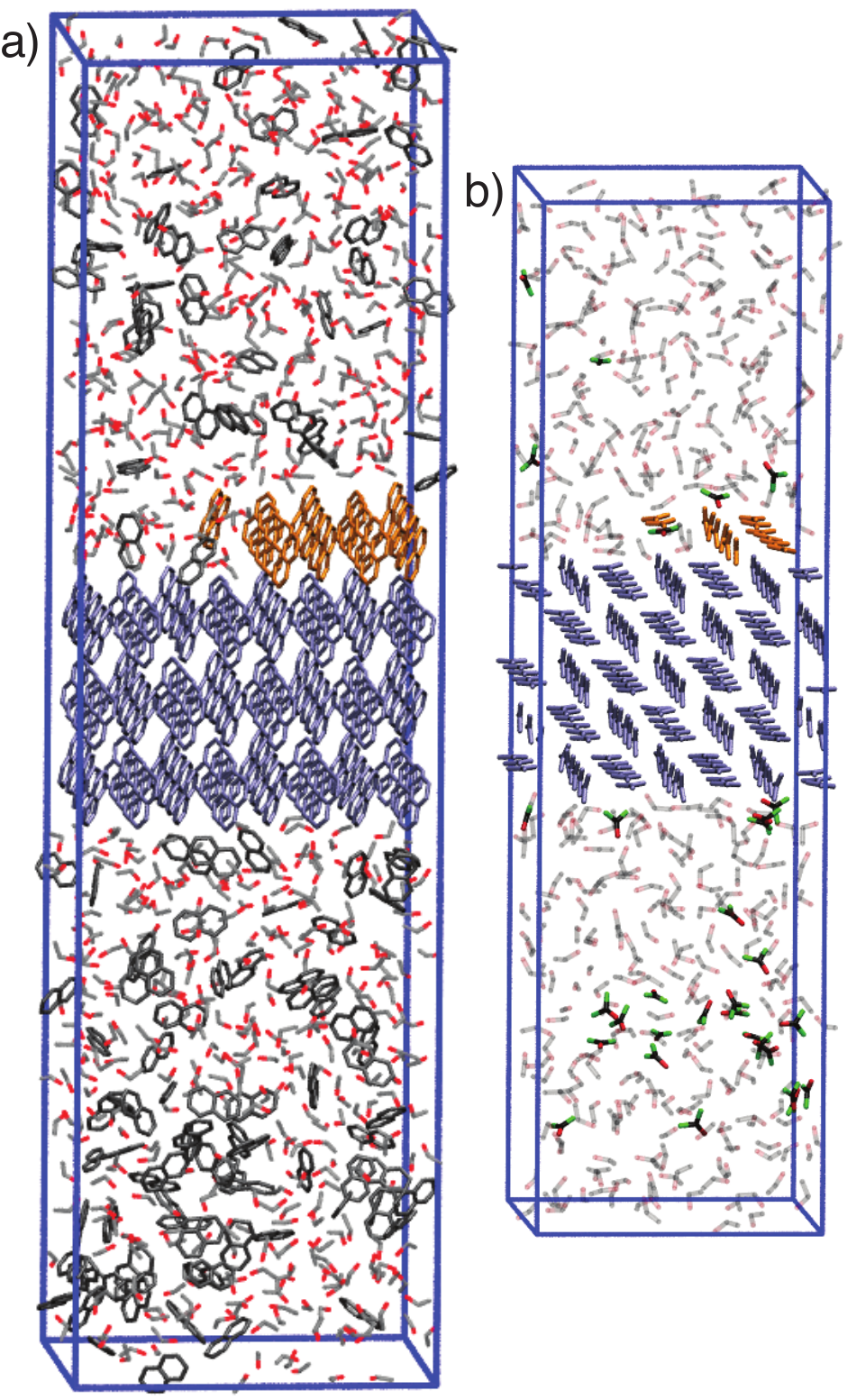}
\end{center}
\caption{Visualizations of the kink growth simulation setups for a) naphthalene grown in ethanol and b) urea grown in ethanol. The biased kink site is located in the center of the upper surface layer. The bulk crystal molecules are colored in blue, the unfinished layer molecules are colored in orange. The atoms of the molecules in solution are colored in black for carbon, red for oxygen and green for nitrogen. Hydrogens are omitted for clarity. Ethanol molecules are shown in faded colors.} \label{fig:setup}
\end{figure}

Kink growth simulations were performed for urea polymorph I \cite{Sklar1961} grown from acetonitrile-, ethanol-, and methanol solutions, as well as for naphthalene polymorph I \cite{Cruickshank1957} grown from ethanol- and toluene solutions. Urea simulations were run at 300 K while naphthalene simulations were run at 280 K, in accordance with our previous work \cite{Bjelobrk2019}.
For both solutes, the slowest growing face was exposed to the solution, i.e. for urea face $\{110\}$ and for naphthalene face $\{00\bar{1}\}$. The unfinished layer was cut along the slowest growing edge, i.e. for urea along the $[0 0 1]$ direction and for naphthalene along the $[0 1 0]$ direction. For each simulation setup, one edge of the surface comprises an unfinished row and the biased kink site. The simulation box equilibration procedure\cite{Parrinello1981} to obtain the initial configurations is reported in the SI.
The visualizations \cite{Humphrey1996} of the simulation setups are shown in Figure \ref{fig:setup} for the cases of naphthalene and urea, each grown from an ethanol solution.

Small displacements of the crystal can occur during the simulations, which can affect the localization protocol of the biased kink site. To avoid such displacements, the movement of the bulk crystal's center of mass was constrained in all three spatial directions by means of a harmonic potential. 

All simulations were run for at least 750 ns. The solution concentration was kept constant with the C$\mu$MD algorithm \cite{Perego2015,Karmakar2018,Karmakar2019,Han2019,Bjelobrk2019}. The CV $s_\text{b}$ was biased through WTMetaD. The WTMetaD Gaussian bias deposition was switched off after 300 ns to increase the convergence performance. All parameter values and further details can be found in the SI.

\section{Results}

To compute the solubilities of the different systems we employed WTMetaD to accelerate the growth and dissolution of the solute molecules from kink sites. Using WTMetaD we sample the free energy difference between the grown and dissolved states, $\Delta F$, thereby allowing us to obtain the solubility values corresponding to the solute mole fraction of the system when $\Delta F = 0$.

\begin{figure}[!htbp]
\begin{center}
\includegraphics[width=\textwidth]{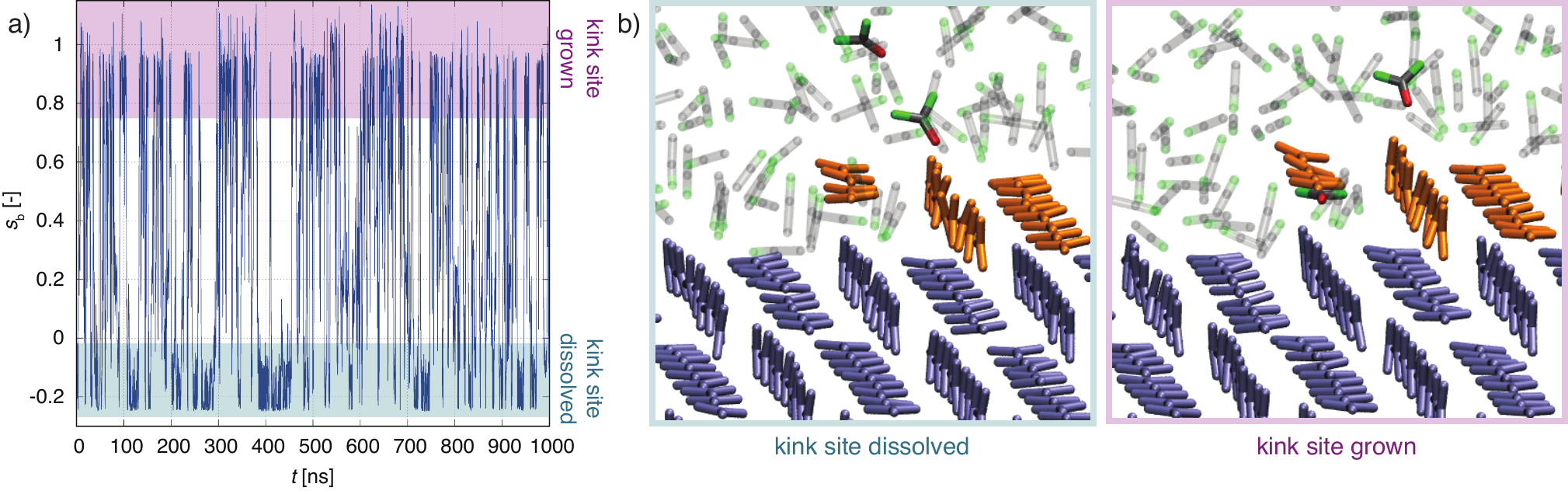}
\end{center}
\caption{a) Biased CV, $s_\text{b}$, in dependence of time, $t$, for urea grown from acetonitrile at a mole fraction of $x = 0.0018$. The regions of the dissolved and grown kink site are shaded in blue and violet respectively. Since we are using negative numbers for weights of the solvent adsorption/desorption, $s_\text{b}$ exhibits negative values for the dissolved kink site states. b) Representative visualizations of the dissolved (left) and grown (right) kink site. The biased kink site is located in the center of each frame. The same color code was used as in Figure \ref{fig:setup}. Hydrogens are omitted for clarity. Acetonitrile molecules are shown in faded colors.} \label{fig:biasplot}
\end{figure}

A typical time evolution of the biased CV, $s_\text{b}$, is shown in Figure \ref{fig:biasplot} for the case of urea face $\{110\}$ exposed to an acetonitrile solution at a solute mole fraction of $x = 0.0018$. One can clearly see that through the WTMetaD bias, many growth and dissolution events at the kink site are obtained for a simulation time of 1 $\mu$s. This is in sharp contrast to an unbiased simulation, where only a few growth or dissolution events would be observed within the same simulation time interval. Figure \ref{fig:biasplot}b) shows representative visualizations of the dissolved and grown kink site states.

\begin{figure}[!htbp]
\begin{center}
\includegraphics[width=15cm]{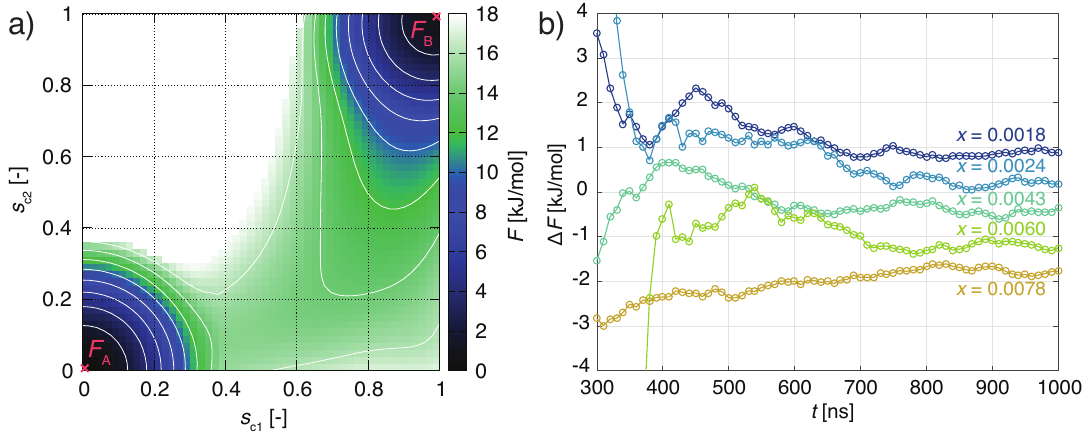}
\end{center}
\caption{a) Free energy surface in dependence of the crystallinity CVs, $s_\text{c1}$ and $s_\text{c2}$, for the case of urea grown from acetonitrile solution at a mole fraction of $x = 0.0018$. The difference in energy of the grown kink site, $F_\text{B}$, and dissolved kink site, $F_\text{A}$, provides $\Delta F$. b) Time evolution of $\Delta F$ for five different mole fractions of urea in acetonitrile.} \label{fig:rewplot}
\end{figure}

Although $s_\text{b}$ is a potent CV for enhancing the growth and dissolution of kink sites, it does not capture the crystallinity of the kink site, which is necessary to compute $\Delta F$. Therefore we reweigh the results using the approach of Tiwary et al. \cite{Tiwary2014} with the aforementioned crystallinity CVs, $s_\text{c1}$ and $s_\text{c2}$. For urea, we take the carbon and oxygen atoms as reference for $s_\text{c1}$ and $s_\text{c2}$. 
For naphthalene we take the centers of the two outermost carbon atom pairs along the naphthalene's long axis as reference (see SI for details).
The first 300 ns of the simulation were used to construct the WTMetaD bias potential. After the first 300 ns, the Gaussian bias deposition was stopped to continue the simulation with a static bias. Only the parts of the simulation with a static bias were used for the reweighing. The corresponding reweighed free energy surface at $t = 1$ $\mu$s is shown in Figure \ref{fig:rewplot}a), from which the energy difference between the crystallized and dissolved kink site states, $\Delta F = F_\text{B} - F_\text{A}$, is obtained. Figure \ref{fig:rewplot}b) shows the time evolution of the reweighed $\Delta F$ for five different mole fractions. $\Delta F$ converges in all cases at $t \approx 700$ ns.

\begin{figure}[!htbp]
\begin{center}
\includegraphics[width=15cm]{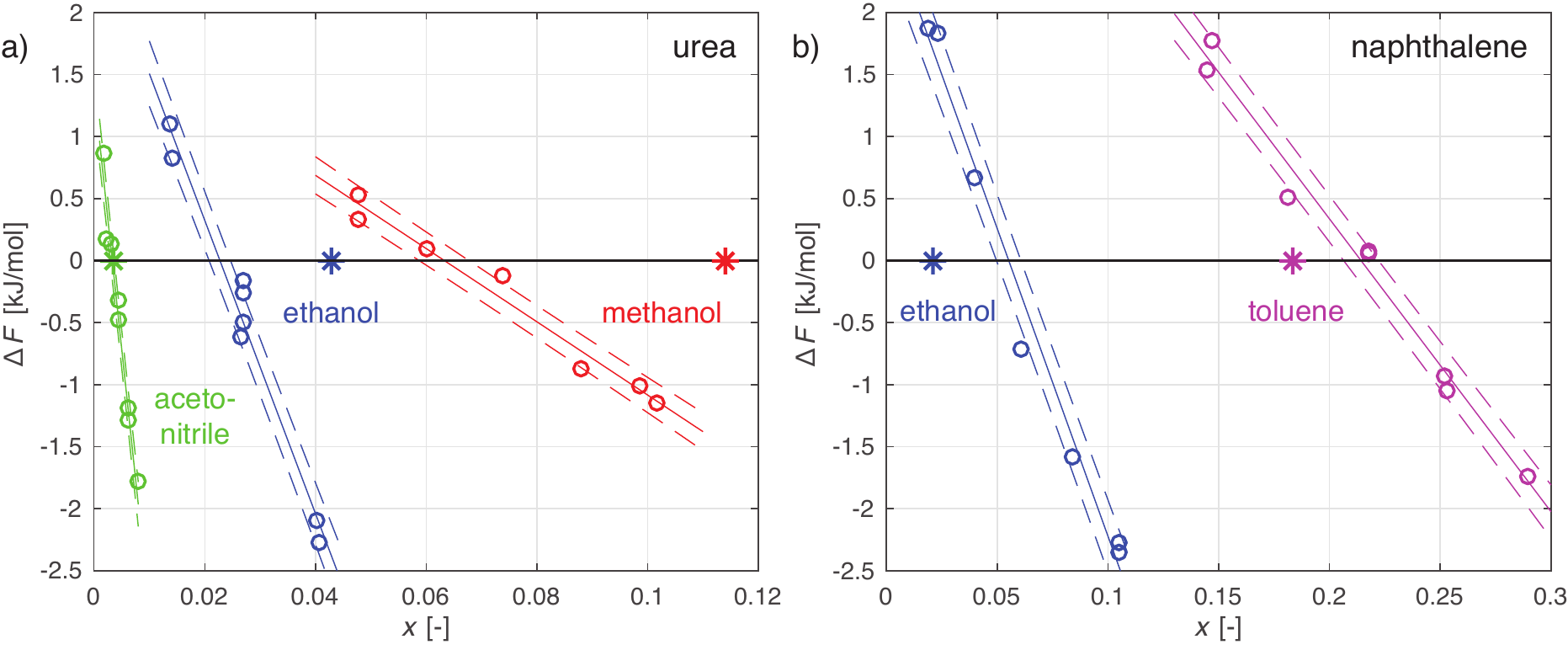}
\end{center}
\caption{Sampled $\Delta F$ in dependence of mole fraction $x$ (circles) in comparison to experimental solubilities (asterisks); a) results for urea grown from acetonitrile (green), ethanol (blue), and methanol (red); b) naphthalene grown from ethanol (blue) and toluene (violet). The straight lines correspond to linear fits of the sampled $\Delta F$ and the dashed lines are the corresponding standard deviations. The simulated solubilities are obtained from interpolation of the linear fit at $\Delta F = 0$.} \label{fig:results}
\end{figure}

From the sampling of $\Delta F$ for different mole fractions one can obtain the predicted solubility, $x^*_\text{sim}$, through interpolation. The graphs in Figure \ref{fig:results} show the results for all sampled $\Delta F$ series of urea grown in acetonitrile, ethanol, and methanol and naphthalene grown in ethanol and toluene (circles). The corresponding experimental solubilities reported in the literature \cite{Loeser2011, Lee1972, Speyers1902, Ward1926} are also shown for comparison (asterisks). 
We approximate the correlation between $\Delta F$ and $x$ as linear within the sampled mole fraction ranges. The graphs show the linear regression for each solvent's $\Delta F$ series together with the lower and upper bound determined by the corresponding standard deviation. We locate $x^*_\text{sim}$ where the linear regression intersects the horizontal axis, i.e. $\Delta F = 0$. The numerical values of the predicted solubility values are listed in Table \ref{tab:results} together with the experimental values.

The comparison between simulations and experiments clearly shows that in all cases the simulations predict the order of magnitude of the values of solubility correctly. Moreover, for both urea and naphthalene the solubility trends in the different solvents are predicted correctly as well. 

More specifically for urea, the solubility in ethanol and methanol is underestimated, whilst that in acetonitrile is predicted exactly. We know from previous studies that the calculated force field melting point, $T_\text{sim}^\text{m} \approx 420$ K, is above its experimental counterpart, $T_\text{exp}^\text{m} = 406$ K \cite{Giberti2015, Salvalaglio2012}. This indicates that the cohesive energy between the urea molecules described by the empirical potential is larger than the actual one and it further implies that the release of urea molecules from the crystal surface is less easy than in reality.
The solubility of urea in the alcohols is therefore underestimated.
Nevertheless, in the case of the urea solubility in acetonitrile, this effect is compensated by the force fields' overemphasis of the solute-solvent interactions through the amine (urea) and nitrile (acetonitrile) functional groups.

Contrariwise, in the case of naphthalene, the force field's calculated melting point, $T_\text{sim}^\text{m} \approx 328$ K, is below the experimental one, $T_\text{exp}^\text{m} = 353$ K. Thus, for naphthalene the lower cohesive force results in facile dissolution of crystalline naphthalene molecules in the solvents and thereby in an overestimated solubility.
The large deviation of solubility of naphthalene in ethanol is due to the force field's underemphasis of solute-solute interactions.
While in the case of toluene, the smaller discrepancy in solubility between simulation and experiment can be anticipated based on the underestimation of the solute-solvent interactions mediated by the underemphasis of the $\pi - \pi$ stacking interactions between their aromatic moieties.

For urea and naphthalene it suffices to sample only one kink site to obtain the solubility, since the chemical environment upon the growth of one molecule along the chosen edge does not change in terms of crystal surface energy. For molecules with more complex growth units such as dimers one has to sample the growth and dissolution of each sub-unit at the particular kink site to obtain the solubility by averaging the energy difference between dissolved and crystalline states over all these sites\cite{Chernov1998, Kuvadia2011}.

It is important to underline, that the simulation method used in this work allowed us to obtain converged values of $\Delta F$ for mole fractions as low as $x = 0.0018$, i.e. corresponding to a mass fraction of $w = 0.006$, as demonstrated for the case of urea in acetonitrile. As a consequence, this simulation setup can extract the solubility in the relevant concentration range for small real life drug intermediates and APIs. For combinations of solute and solvent, where the solubility is lower than a mass fraction of roughly $w = 0.01$, such setup cannot be used to estimate solubility. It is also worth noting that in these cases the species is considered to be insoluble as far as the design of a crystallization process in the pharmaceutical industry is concerned.

\begin{table}
\centering
\captionsetup{justification=centering}
\caption{Predicted solubilities, $x_\text{sim}^*$, compared to experimental solubilities, $x_\text{exp}^*$. The \\simulations were performed at 1 bar and 300 K for urea, and 280 K for naphthalene.}
\begin{tabular}{llllllll}
\toprule
                  & & \multicolumn{3}{c}{urea} & & \multicolumn{2}{c}{naphthalene} \\
                  & & MeCN       & EtOH        & MeOH        & & EtOH   & MePh            \\
\midrule
$x_\text{sim}^*$  & & 0.0033(4)  & 0.0228(20)  & 0.0633(45)  & & 0.0552(55)  & 0.2143(79)  \\
$x_\text{exp}^*$ \cite{Loeser2011, Lee1972, Speyers1902, Ward1926}
                  & & 0.0037     & 0.0429      & 0.1140      & & 0.0213     & 0.183       \\
\bottomrule
\end{tabular}
\label{tab:results}
\end{table}

\section{Conclusions}

We have introduced a new approach to predict the solubility of organic molecular crystals in a variety of solvents using molecular dynamics. The approach samples energy differences between grown and dissolved kink site states at controlled solution concentrations \cite{Perego2015}. To achieve sufficient sampling performance of the growth and dissolution events, well-tempered Metadynamics was used through a collective variable that has been developed as a part of this work. The collective variable addresses all relevant slow degrees of freedom of the kink growth process, namely the solute diffusion to and adsorption/desorption at the kink site as well as the solvent desorption/adsorption at the kink site.

We have shown that using General Amber Force Fields for urea crystals grown in acetonitrile, ethanol and methanol solutions and for naphthalene crystals grown in ethanol and toluene solutions, the computed solubility values exhibit the correct order of magnitude and the correct trends in the different solvents when compared with experimental measurements. The deviation of the melting point of the force field from its experimental counterpart allows us to predict whether in the simulations the solubility will be rather underestimated, as for urea, whose force field's melting temperature is too high, or overestimated, as for naphthalene, whose force field's melting temperature is too low in comparison to experiments.

Naturally, here we obtain the solubility of the force field. However with the rapid advance of machine learning based force fields, which can represent the atomistic properties to \textit{ab initio} accuracy \cite{Bonati2018}, it is reasonable to expect that in the near future the presented method will be able to compute solubility significantly closer to its experimental counterpart.

We anticipate that an extension of the setup used here will allow also the calculation of the activation energies of kink growth, which can be important for reliable parameter estimation of spiral growth models.

\section*{Acknowledgements}
Z. B. and M. M. thank Novartis Pharma AG for their partial financial support to this project.
Z. B. thanks Pablo Piaggi, Ashwin Rajagopalan, Michele Invernizzi, Thilo Weber, Philipp Müller, and Marco Holzer for valuable discussions.
The computational resources were provided by ETH Zürich and the Swiss Center for Scientific Computing at the Euler Cluster.

\bibliographystyle{unsrt} 


\bibliography{bibliography} 

\newpage
\begin{center}
\vspace{5mm}
\LARGE{\textsc{Supporting information}}
\vspace{2mm}
\end{center}

\section*{Simulation setup equilibration}

For all equilibration simulations we used Gromacs 2016.5 \cite{Abraham2015} with full atomistic description of all molecule types.
We used the velocity rescaling thermostat \cite{Bussi2009}, periodic boundary conditions, the Ewald particle mesh approach \cite{Darden1993} for the electrostatic interactions, and the LINCS algorithm \cite{Hess2008a,Hess2008b} to constrain the covalent bonds involving hydrogens. The non-bonded cutoff was set to 1 nm.

The simulation box specifications including the box lengths $L_x$, $L_y$, and $L_z$, as well as number of solute molecules $N_s$ and solvent molecules $N_l$ are listed in Table \ref{tab:boxspecs} for all studied systems in this work, namely urea grown in acetonitrile (MeCN), ethanol (EtOH), and methanol (MeOH), and naphthalene grown in ethanol and toluene (MePh). For the urea-MeCN system, a larger simulation box was used to reach the low solubility regime. The solubility does not depend on the simulation box sizes for the reported setups. However, simulation box sizes which are smaller than the reported ones do suffer from finite size effects, especially if the length of the liquid phase in $z$ direction (perpendicular to the crystal surface) is not long enough. Too short liquid phase lengths can lead to a weak orientation pattern of the solvent which can cause a significant drop in solubility for the given solute compound.

\begin{table}[!htbp]
\caption{Simulation box specifications of all studied systems.}
\centering
\begin{tabular}{lcccccccc}
\toprule
                   & & \multicolumn{3}{c}{urea} & & \multicolumn{2}{c}{naphthalene} \\
                   & & MeCN     & EtOH     & MeOH     & & EtOH     & MePh           \\
\midrule
$N_\text{s}$ [-]   & &      504 &      225 &      250 & &      260 &      320       \\
$N_\text{l}$ [-]   & &     1400 &      390 &      525 & &      706 &      350       \\
$L_x$ [nm]         & &  3.78200 &  2.26842 &  2.26842 & &  3.28663 &  3.28663       \\
$L_y$ [nm]         & &  4.25456 &  2.83593 &  2.83593 & &  2.98478 &  2.98478       \\
$L_z$ [nm]         & & 10.46135 &  8.22554 &  8.06331 & & 12.13055 & 12.96988       \\
$T$ [K]            & &      300 &      300 &      300 & &      280 &      280       \\  
$p$ [bar]          & &        1 &        1 &        1 & &        1 &        1       \\
\bottomrule
\end{tabular}
\label{tab:boxspecs}
\end{table}

It suffices to perform simulations under $NVT$ instead of the computationally more expensive $NPT$ conditions, since the growth of a single kink site does not noticeably alter the pressure of the system. To obtain the appropriate simulation box lengths we used the following equilibration protocol for all considered systems.

First, a seed crystal was constructed from XRD data (urea polymorph I \cite{Sklar1961} and naphthalene polymorph I \cite{Cruickshank1957}) with the face of interest perpendicular to the $z$-axis: for urea face $\{110\}$ and for naphthalene face $\{00\bar{1}\}$. The crystal system energy was then minimized with the conjugate gradient algorithm with a tolerance of the maximum force of 50 kJ mol$^{-1}$ nm$^{-1}$, followed by a temperature equilibration at $NVT$ conditions for 1 ns at an integration time step of 0.5 fs, to reach the targeted temperatures. The pressure equilibration was achieved by running the simulation setup for a further 25 ns at $NPT$ conditions using the anisotropic Parrinello-Rahman barostat \cite{Parrinello1981} with the same integration time step of 0.5 fs. The data of the last 20 ns of the simulation were used to calculate the average box lengths $L_x$, $L_y$,  and their average ratio $L_x/L_y$ to identify the simulation box frame closest to the average values.

Second, we submerged each of the crystals in the corresponding solute-solvent mixture using the genbox utility of Gromacs \cite{Hess2008b}. The same energy minimization, and $NVT$ equilibration were performed as for the crystal equilibration step. 
For the $NPT$ equilibration we used the semi-isotropic barostat to allow expansion/contraction of the simulation box only along the $z$-axis while keeping the already averaged $L_x$ and $L_y$ constant.
From the $NPT$ equilibration run we used the last 20 of the 25 ns to compute the average box length $L_z$ at the pressure of 1 bar. Again, the simulation frame with the simulation box length along $z$ closest to $L_z$ was chosen as initial configuration for the $\mu VT$ equilibration step.

Third, we used the C$\mu$MD algorithm \cite{Perego2015} to obtain the targeted concentration profiles in the vicinity of the crystal surface. A simulation time of 25 ns ensures to reach the targeted concentration profile.
During the $\mu VT$ equilibration, a harmonic potential was used to push the molecules away from the crystal surface, which do not belong to the unfinished layer. The unfinished surface layer was prevented from dissolving using a potential acting through the surface structure CV.

\begin{figure}[!htbp]
\begin{center}
\includegraphics[width=\textwidth]{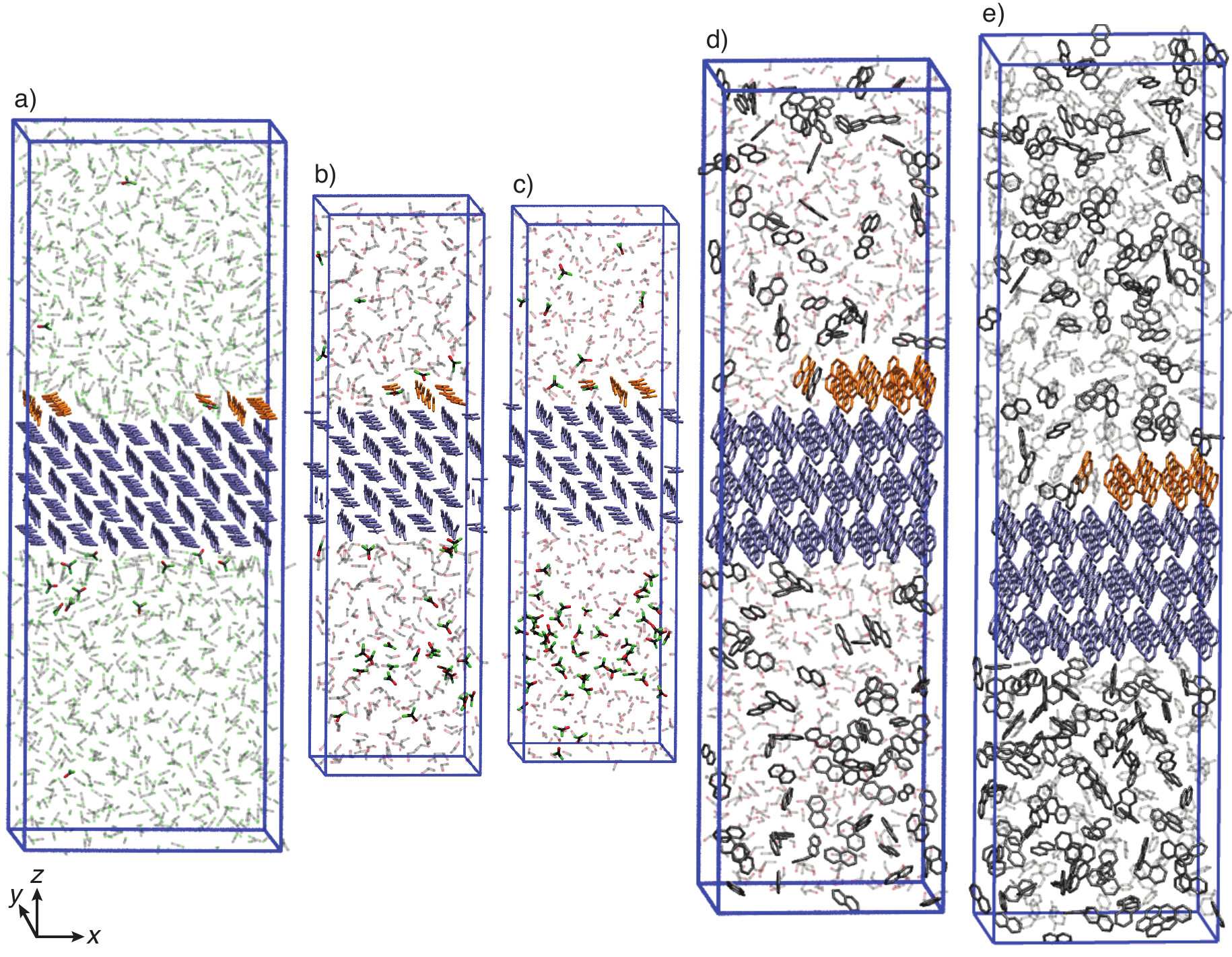}
\end{center}
\caption{Simulation box visualizations; a) urea in acetonitrile, b) urea in ethanol, c) urea in methanol, d) naphthalene in ethanol, e) naphthalene in toluene. The bulk crystal molecules are colored in blue, the unfinished surface layer molecules are colored in orange. The atoms of the molecules in solution are colored in black for carbon, red for oxygen and green for nitrogen. Hydrogens are omitted for clarity. Solvent molecules are shown in faded colors.} \label{fig:simulationboxes}
\end{figure}

\begin{figure}[!htbp]
\begin{center}
\includegraphics[width=\textwidth]{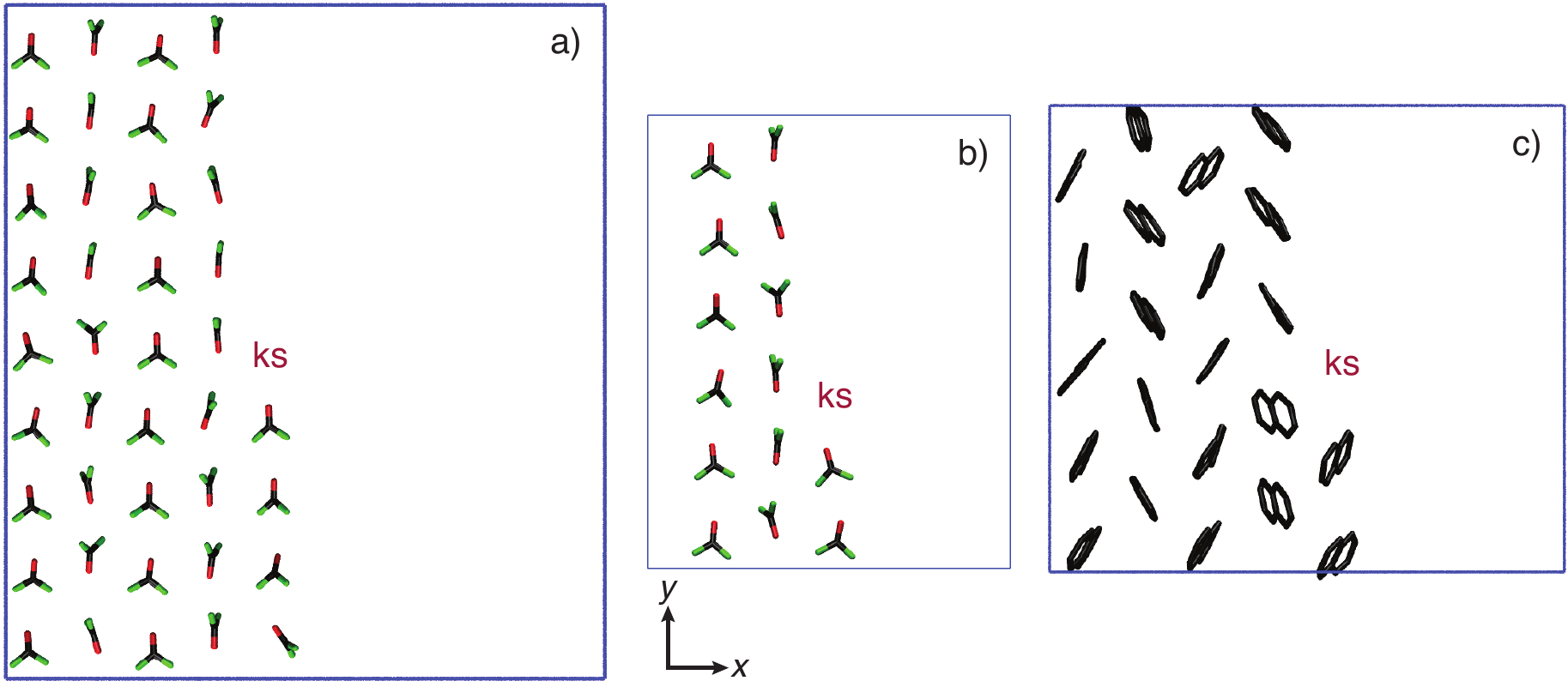}
\end{center}
\caption{Visualizations of the unfinished surface layer molecules with the location of the biased kink site indicated by 'ks'. The atoms of the molecules are colored in black for carbon, red for oxygen and green for nitrogen. Hydrogens are omitted for clarity.} \label{fig:surfacelayers}
\end{figure}

Simulation box visualizations are shown in Figure \ref{fig:simulationboxes} and the unfinished surface layer visualizations are shown in Figure \ref{fig:surfacelayers}. The surface layer was cut along the $[0 0 1]$ direction for urea and along the $[0 1 0]$ direction for naphthalene.

\section*{Constant chemical potential method}\label{sec:CmuMD}

\begin{SCfigure}[0.5][h]
\includegraphics[width=0.5\textwidth]{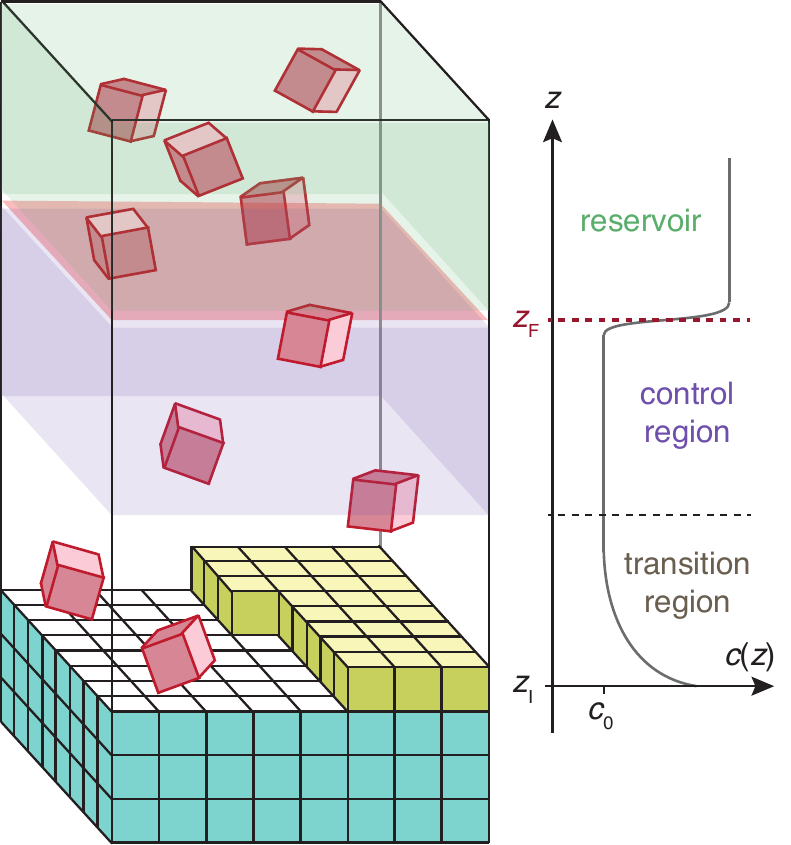}
\caption{Scheme of the simulation setup with corresponding concentration profile, $c(z)$, along the $z$ axis. The solute molecules are shown as cubes where the bulk crystal molecules are colored in green, the molecules of the unfinished layer, which comprises the kink site, are colored in yellow and the dissolved molecules are colored in red. Depiction of solvent molecules is omitted for clarity. The control region is shaded in violet and the reservoir is shaded in green. The position, $z_\text{F}$, where the external force acts is colored in red.} \label{fig:CmuMD}
\end{SCfigure}

We briefly discuss the C$\mu$MD method \cite{Perego2015}. The scheme shown in Figure \ref{fig:CmuMD} depicts the simulation setup, with the corresponding solute concentration profile, $c(z)$, along the $z$ axis. Periodic boundary conditions are introduced for all spatial directions. The crystal surface, with an unfinished layer comprising the kink site, is exposed to the solution.

To keep the solution concentration constant the C$\mu$MD algorithm is introduced, which works as follows. The liquid phase of the simulation box is partitioned along the $z$ axis into following segments: a transition region, control region and reservoir. An external force $F^\mu_i$ is introduced to control the flux of solutes $i$ between the control region and reservoir at position $z_\text{F}$. $F^\mu_i$ is defined as follows:
\begin{equation}
F^\mu_i = k^\mu (c_\text{CR}(t) - c_0) G_\omega(z_i,z_\text{F}).
\end{equation}
$k^\mu$ is a force constant, $c_\text{CR}(t)$ is the concentration of the control region at time $t$, and $c_0$ is the predefined target concentration. $G_\omega(z_i,z_\text{F})$ is a bell shaped function:
\begin{equation}
G_\omega(z_i,z_\text{F}) = \frac{1}{4\omega} \left[ 1 + \cosh \left( \frac{z_i - z_\text{F}}{\omega} \right) \right]^{-1},
\end{equation}
where $z_i$ corresponds to the $z$ position of solute molecule $i$ and $\omega$ defines the height and width of the bell curve.

If $c(t)$ is at a given time step below $c_0$, then $F^\mu_i$ will accelerate the solute molecules from the reservoir towards the control region and vice versa. This creates a constant concentration profile in the control region, as shown in Figure \ref{fig:CmuMD}, and enables the simulation of kink growth at constant chemical potential. See ref. \citenum{Perego2015} for further details on the C$\mu$MD method.

The C$\mu$MD parameters used in the simulations are listed in Table \ref{tab:CmuMD}.

\begin{table}[!htbp]
\caption{Values of the C$\mu$MD parameters.}
\centering
\begin{tabular}{rcccccccc}
\toprule
                      & & \multicolumn{3}{c}{urea} & & \multicolumn{2}{c}{naphthalene} \\
                      & & MeCN  & EtOH  & MeOH  & & EtOH  & MePh  \\
             
\midrule
$\omega/L_z$ [-]      & & 0.02  & 0.02  & 0.02  & & 0.02  & 0.02  \\
$z_\text{TR}/L_z$ [-] & & 0.16  & 0.13  & 0.22  & & 0.14  & 0.24  \\
$z_\text{CR}/L_z$ [-] & & 0.50  & 0.20  & 0.22  & & 0.18  & 0.20  \\
$z_\text{F}/L_z$ [-]  & & 0.68  & 0.37  & 0.50  & & 0.34  & 0.48  \\
$\Delta z/L_z$ [-]    & & 1/120 & 1/120 & 1/120 & & 1/120 & 1/120 \\
\bottomrule
\end{tabular}
\label{tab:CmuMD}
\end{table}

\section*{Collective variables (CVs)}

\subsection*{Biased CV}

The values used for the biased CV $s_\text{b}$ are listed in Table \ref{tab:biasedCV}.

\begin{table}
\caption{Values of the biased CV used in the simulation.}
\centering
\begin{tabular}{cccccccccccc}
\toprule
&                         & & \multicolumn{3}{c}{urea} & & \multicolumn{2}{c}{naphthalene} \\
&                         & & MeCN   & EtOH   & MeOH   & & EtOH   & MePh   \\
\midrule
\multirow{7}{*}{$s_\text{s}$}
& $r_\text{s}^{(x)}$ [nm] & & 2.8366 & 1.1364 & 1.1384 & & 1.8129 & 3.0464 \\
& $r_\text{s}^{(y)}$ [nm] & & 2.2038 & 1.1096 & 1.1091 & & 1.3490 & 1.3464 \\
& $r_\text{s}^{(z)}$ [nm] & & 1.1404 & 1.3231 & 1.4044 & & 1.4945 & 0.1121 \\
& $\sigma_\text{s}$ [-]   & & 0.30   & 0.20   & 0.20   & & 0.4    & 0.22   \\
& $\chi_\text{s1}$ [-]    & & 0.3    & 0.3    & 0.3    & & 0.3    & 0.3    \\
& $\chi_\text{s2}$ [-]    & & 3      & 3      & 3      & & 3      & 3      \\
& $w_\text{s}$ [-]        & & 0.5    & 0.5    & 0.5    & & 0.5    & 0.5    \\ \midrule
\multirow{6}{*}{$s_\text{l}$}
& $r_\text{l}^{(x)}$ [nm] & & 2.8366 & 1.1364 & 1.1384 & & 1.8129 & 3.0464 \\
& $r_\text{l}^{(y)}$ [nm] & & 2.2038 & 1.1096 & 1.1091 & & 1.3490 & 1.3464 \\
& $r_\text{l}^{(z)}$ [nm] & & 1.1404 & 1.3231 & 1.4044 & & 1.4945 & 0.1121 \\
& $\sigma_\text{l}$ [-]   & & 0.1    & 0.1    & 0.1    & & 0.2    & 0.2    \\
& $\chi_\text{l}$ [-]     & & 0.3    & 0.3    & 0.3    & & 0.5    & 0.5    \\
& $w_\text{l}$ [-]        & & $-$0.25  & $-$0.25  & $-$0.25  & & $-$0.25  & $-$0.5   \\
\bottomrule
\end{tabular}
\label{tab:biasedCV}
\end{table}

To improve the sampling performance of the WTMetaD simulations, lower and upper wall potentials were used for the biased CV:
\begin{equation}
V_\text{s} =
\begin{cases}
k_\text{s,l} ((s_\text{s}^{\chi_\text{s1}}+s_\text{s}^{\chi_\text{s2}}) - s_\text{s,l})^2, & \text{if}\ (s_\text{s}^{\chi_\text{s1}}+s_\text{s}^{\chi_\text{s2}}) < s_\text{s,l}, \\
k_\text{s,u} ((s_\text{s}^{\chi_\text{s1}}+s_\text{s}^{\chi_\text{s2}}) - s_\text{s,u})^2, & \text{if}\ (s_\text{s}^{\chi_\text{s1}}+s_\text{s}^{\chi_\text{s2}}) > s_\text{s,u}, \\
0, & \text{else},
\end{cases}
\quad \text{and} \quad
V_\text{l} =
\begin{cases}
k_\text{l,l} (s_\text{l}^{\chi_\text{l}} - s_\text{l,l})^2, & \text{if}\ s_\text{l}^{\chi_\text{l}} < s_\text{l,l}, \\
k_\text{l,u} (s_\text{l}^{\chi_\text{l}} - s_\text{l,u})^2, & \text{if}\ s_\text{l}^{\chi_\text{l}} > s_\text{l,u}, \\
0, & \text{else}.
\end{cases}
\end{equation}

where $k_\text{s,l}$, $k_\text{s,u}$, $k_\text{l,l}$, and $k_\text{l,u}$ are the force constants and $s_\text{s,l}$, $s_\text{s,u}$, $s_\text{l,l}$, and $s_\text{l,u}$ are the thresholds below and above which the potentials are acting. The lower walls inhibits the biased simulations from getting stuck at $s_\text{s}$ and $s_\text{l}$ values of zero and the higher walls inhibit excessive agglomeration of solute or solvent molecules at the kink site, which is not relevant for the kink growth process. The values of the potentials used in this work can be found in Table \ref{tab:biaswalls}.

The WTMetaD \cite{Barducci2008} parameter values are presented in Table \ref{tab:MetaDparams}. $W$ and $\sigma_W$ are the height and width of the Gaussians, $\gamma$ the bias factor, $\tau$ the bias deposition stride, and $\Delta s_\text{b}$ is the bin length of the grid on which the bias is stored.

\begin{table}
\caption{Values of the wall potentials parameters used for the biased CV.}
\centering
\begin{tabular}{cccccccccccc}
\toprule
&                         & & \multicolumn{3}{c}{urea} & & \multicolumn{2}{c}{naphthalene} \\
&                         & & MeCN   & EtOH   & MeOH   & & EtOH   & MePh   \\
\midrule
\multirow{4}{*}{$s_\text{s}$}
& $k_\text{s,l}$ [kJ/mol] & & 15     & 15     & 15     & & 15     & 15     \\
& $k_\text{s,u}$ [kJ/mol] & & 15     & 15     & 15     & & 15     & 15     \\
& $s_\text{s,l}$ [-]      & & 0.03   & 0.03   & 0.03   & & 0.05   & 0.04   \\
& $s_\text{s,u}$ [-]      & & 2.08   & 2.03   & 2.03   & & 2.22   & 2.40   \\ \midrule
\multirow{4}{*}{$s_\text{l}$}
& $k_\text{l,l}$ [kJ/mol] & & 15     & 15     & 15     & & 15     & 15     \\
& $k_\text{l,u}$ [kJ/mol] & & 15     & 15     & 15     & & 15     & 15     \\
& $s_\text{l,l}$ [-]      & & 0.025  & 0.025  & 0.025  & & 0.04   & 0.08   \\
& $s_\text{l,u}$ [-]      & & 0.995  & 0.995  & 0.995  & & 1.10   & 1.10   \\
\bottomrule
\end{tabular}
\label{tab:biaswalls}
\end{table}

\begin{table}
\caption{Values of the well-tempered Metadynamics parameters.}
\centering
\begin{tabular}{cccccccc}
\toprule
                        & & \multicolumn{3}{c}{urea} & & \multicolumn{2}{c}{naphthalene} \\
                        & & MeCN  & EtOH  & MeOH  & & EtOH  & MePh  \\
             
\midrule
$W$ [kJ/mol]            & & 0.2   & 0.2   & 0.2   & & 0.2   & 0.4   \\
$\sigma_W$ [-]          & & 0.06  & 0.03  & 0.03  & & 0.03  & 0.03  \\
$\gamma$ [-]            & & 4     & 3     & 3     & & 2     & 4     \\
$\tau$ [ps]             & & 1     & 1     & 1     & & 1     & 1     \\
$\Delta s_\text{b}$ [-] & & 0.02  & 0.01  & 0.01  & & 0.01  & 0.01  \\
\bottomrule
\end{tabular}
\label{tab:MetaDparams}
\end{table}

\subsection*{Surface structure CV}

To prevent the dissolution of the unfinished layer, a harmonic potential wall is introduced through the surface structure CV, $s_\text{st}$. $s_\text{st}$ is defined as the logistic function: 
\begin{equation*}
s_\text{st} = \frac{1}{1 + \exp{(-\sigma_\text{st}(\tilde{s}_\text{st}-\tilde{s}_\text{st,0}))}}, 
\end{equation*}\label{eq:s_st}
with step position $\tilde{s}_{\text{st},0}$ and steepness $\sigma_\text{st}$, of following function:
\begin{equation*}
\tilde{s}_\text{st} = \sum_i \left( \sum_k \left[\cos^{\eta_x} \left( \frac{\nu_x \pi}{L_x} (x_i - \bar{x}_k) \right) \cos^{\eta_y} \left( \frac{\nu_y \pi}{L_y} (y_i - \bar{y}_k) \right)\right] \exp \left\{ -\frac{(z_i-\bar{z})^2}{2 \sigma_z^2} \right\} \right).
\end{equation*}
$\nu_{x/y}$ corresponds to the number of unit cells along the $x$ and $y$ axes, $L_{x/y}$ is the length of the simulation box in $x/y$ direction, $\bar{x}_k$ and $\bar{y}_k$ are the $k$-th molecule center position in $x$ and $y$ direction within the crystal unit cell. The exponent $\nu_{x/y}$ is a positive even integer and defines the width of the sinusoid peaks. For the $z$ part of $s_\text{st}$, $\bar{z}_i$ defines the position in $z$ direction, and $\sigma_z$ is the width of the Gaussian like curve. The expression is summed over all solute molecules $i$ in the unfinished surface layer.

\begin{figure}[!htbp]
\begin{center}
\includegraphics[width=9cm]{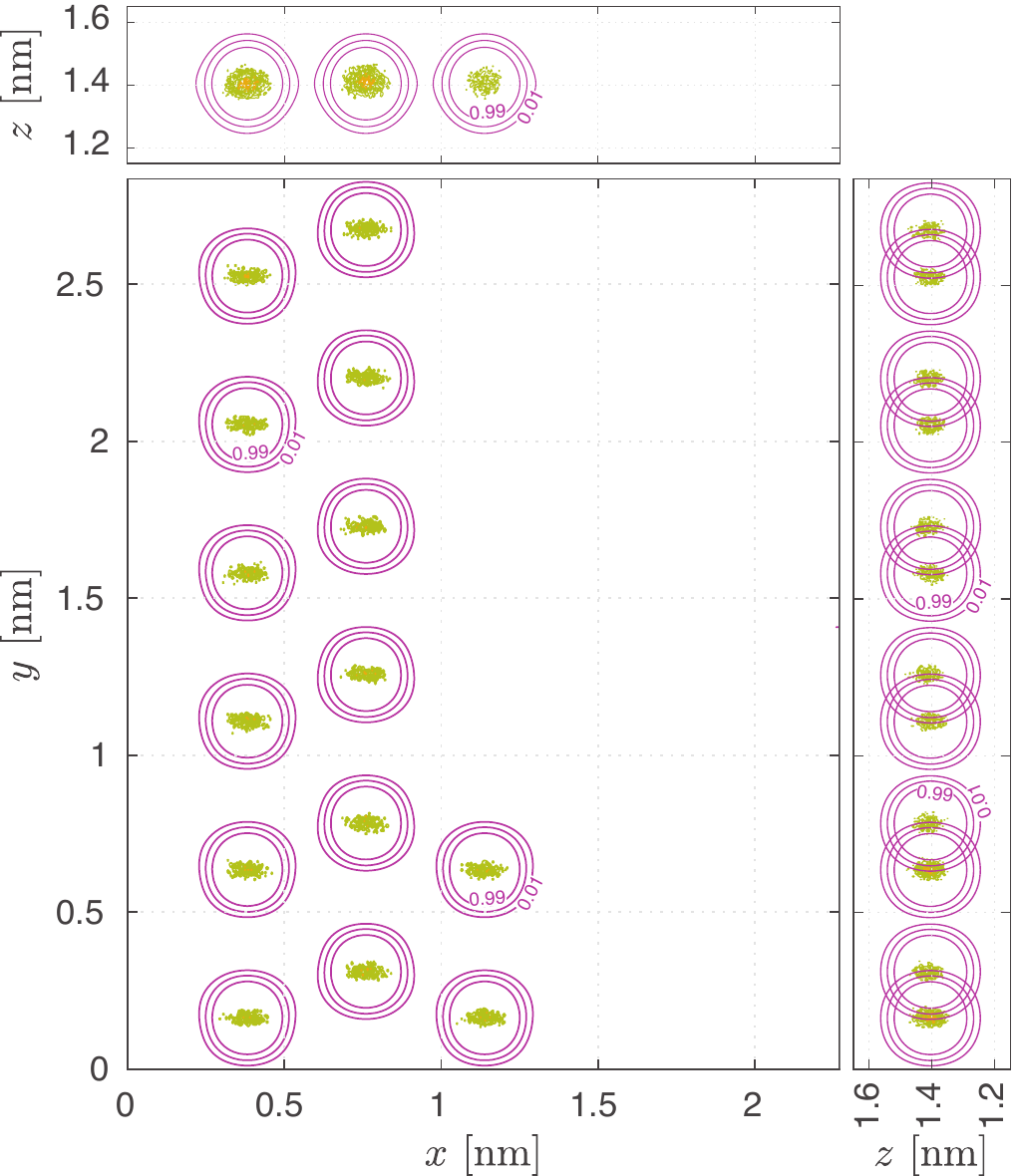}
\end{center}
\caption{Contour lines of the surface structure CV (pink curves) and histogram of urea center positions (yellow dots) of the unfinished surface layer projected along all three spatial directions.} \label{fig:s_st}
\end{figure}

The form of $s_\text{st}$ is such that its value is 1 if the center of the solute molecule is at its adsorption site and otherwise zero. This is achieved by setting the steepness, $\sigma_\text{st}$, and position, $\tilde{s}_\text{st,0}$, of the logistic function step of $s_\text{st}$ accordingly. Figure \ref{fig:s_st} shows the contour lines of $s_\text{st}$ together with the histogram of the urea carbon atom positions of the unfinished layer.

A harmonic wall potential, $V_\text{st}$, is introduced to the system through $s_\text{st}$:
\begin{equation*}
V_\text{st} =
\begin{cases}
k_\text{st} (s_\text{st} - s_\text{st,0})^2, & \text{if}\ s_\text{st} < s_\text{st,0}, \\
0, & \text{else},
\end{cases}
\end{equation*}
where $k_\text{st}$ is the force constant and $s_\text{st,0}$ is the threshold below which the harmonic potential acts. $V_\text{st}$ prevents the surface molecules from dissolving while at the same time it does not interfere with their thermal lattice vibrations as shown in Figure \ref{fig:s_st}. The parameters used in the simulations are shown in Table \ref{tab:s_st}.

While the naphthalene face $\{00\bar{1}\}$ is stable enough that no dissolution of the surface on the opposite site of the crystal or the layer below the unfinished layer is observed within the simulation time spans of $\sim$ 1 $\mu$s, these layers can dissolve for urea face $\{110\}$ grown in ethanol and methanol. We introduced a harmonic potential through the surface structure CV also for the urea layer on the opposite crystal surface (layer 1) and for the layer below the unfinished surface layer (layer 5). The used parameter values are reported in Table \ref{tab:s_st_urea}.

\begin{table}[!htbp]
\caption{Values of the surface structure CV parameters.}
\centering
\begin{tabular}{cccccccc}
\toprule
                         & & \multicolumn{3}{c}{urea}    & & \multicolumn{2}{c}{naphthalene} \\
                         & & MeCN    & EtOH    & MeOH    & & EtOH   & MePh     \\
\midrule
$\nu_x$ [-]            & & 5       & 3       & 3       & & 4       & 4       \\
$L_x$ [nm]               & & 3.78200 & 2.26842 & 2.26842 & & 3.28663 & 3.28663 \\
$\eta_x$ [-]           & & 16      & 16      & 16      & & 16      & 16      \\
$\bar{x}_1$ [nm]         & & -1.3213 & 0.38002 & 0.38002 & & 1.6210  & 1.6210  \\
$\bar{x}_2$ [nm]         & & -0.9434 & 0.75827 & 0.75827 & & 0.3875  & 0.3875  \\
$\nu_y$ [-]            & & 9       & 6       & 6       & & 5       & 5       \\
$L_y$ [nm]               & & 4.25456 & 2.83593 & 2.83593 & & 2.98478 & 2.98478 \\
$\eta_y$ [-]           & & 8       & 8       & 8       & & 8       & 8       \\
$\bar{y}_1$ [nm]         & & -1.4902 & 0.6391  & 0.6391  & & 0.4540  & 0.4540  \\
$\bar{y}_2$ [nm]         & & -1.3425 & 0.7874  & 0.7874  & & 0.1521  & 0.1521  \\
$\bar{z}$ [nm]           & & 0.7538  & 1.2873  & 1.4044  & & 1.4920  & 0.1187  \\
$\sigma_z$ [nm]          & & 0.065   & 0.065   & 0.065   & & 0.65    & 0.65    \\
$\sigma_\text{st}$ [-] & & 150     & 150     & 150     & & 150     & 150     \\
$\tilde{s}_{\text{st},0}$ [-]  & & 0.1     & 0.1     & 0.1     & & 0.1     & 0.1     \\
$k_\text{st}$ [kJ/mol]   & & 15     & 15     & 15     & & 15     & 15     \\
$s_{\text{st},0}$ [-] & & 40     & 14     & 14     & & 22     & 22     \\
\bottomrule
\end{tabular}
\label{tab:s_st}
\end{table}

\begin{table}[!htbp]
\caption{Values of the surface structure CV parameters for the urea bulk crystal layers.}
\centering
\begin{tabular}{ccccccc}
\toprule
                              & & \multicolumn{2}{c}{urea layer 1} & & \multicolumn{2}{c}{urea layer 5} \\
                              & & EtOH      & MeOH      & & EtOH    & MeOH    \\
\midrule
$\nu_x$ [-]                   & & 3         & 3         & & 3       & 3       \\
$L_x$ [nm]                    & & 2.26842   & 2.26842   & & 2.26842 & 2.26842 \\
$\eta_x$ [-]                  & & 16        & 16        & & 16      & 16      \\
$\bar{x}_1$ [nm]              & & 0.3800    & 0.3799    & & 0.3800  & 0.3801  \\
$\bar{x}_2$ [nm]              & & 0.7583    & 0.7583    & & 0.7583  & 0.7582  \\
$\nu_y$ [-]                   & & 6         & 6         & & 6       & 6       \\
$L_y$ [nm]                    & & 2.83593   & 2.83593   & & 2.83593 & 2.83593 \\
$\eta_y$ [-]                  & & 8         & 8         & & 8       & 8       \\
$\bar{y}_1$ [nm]              & & 0.6391    & 0.6376    & & 0.6391  & 0.6376  \\
$\bar{y}_2$ [nm]              & & 0.7874    & 0.7862    & & 0.7874  & 0.7875  \\
$\bar{z}$ [nm]                & & $-$0.6009 & $-$0.4842 & & 0.9091  & 1.0259  \\
$\sigma_z$ [nm]               & & 0.065   & 0.065       & & 0.065   & 0.065   \\
$\sigma_\text{st}$ [-]        & & 150     & 150         & & 150     & 150     \\
$\tilde{s}_{\text{st},0}$ [-] & & 0.1     & 0.1         & & 0.1     & 0.1     \\
$k_\text{st}$ [kJ/mol]        & & 10      & 10          & & 10      & 10      \\
$s_{\text{st},0}$ [-]         & & 36      & 36          & & 36      & 36      \\
\bottomrule
\end{tabular}
\label{tab:s_st_urea}
\end{table}

\subsection*{Adsorption site CVs} \label{sec:adsorptionCVs}

\begin{figure}[!htbp]
\begin{center}
\includegraphics[width=7cm]{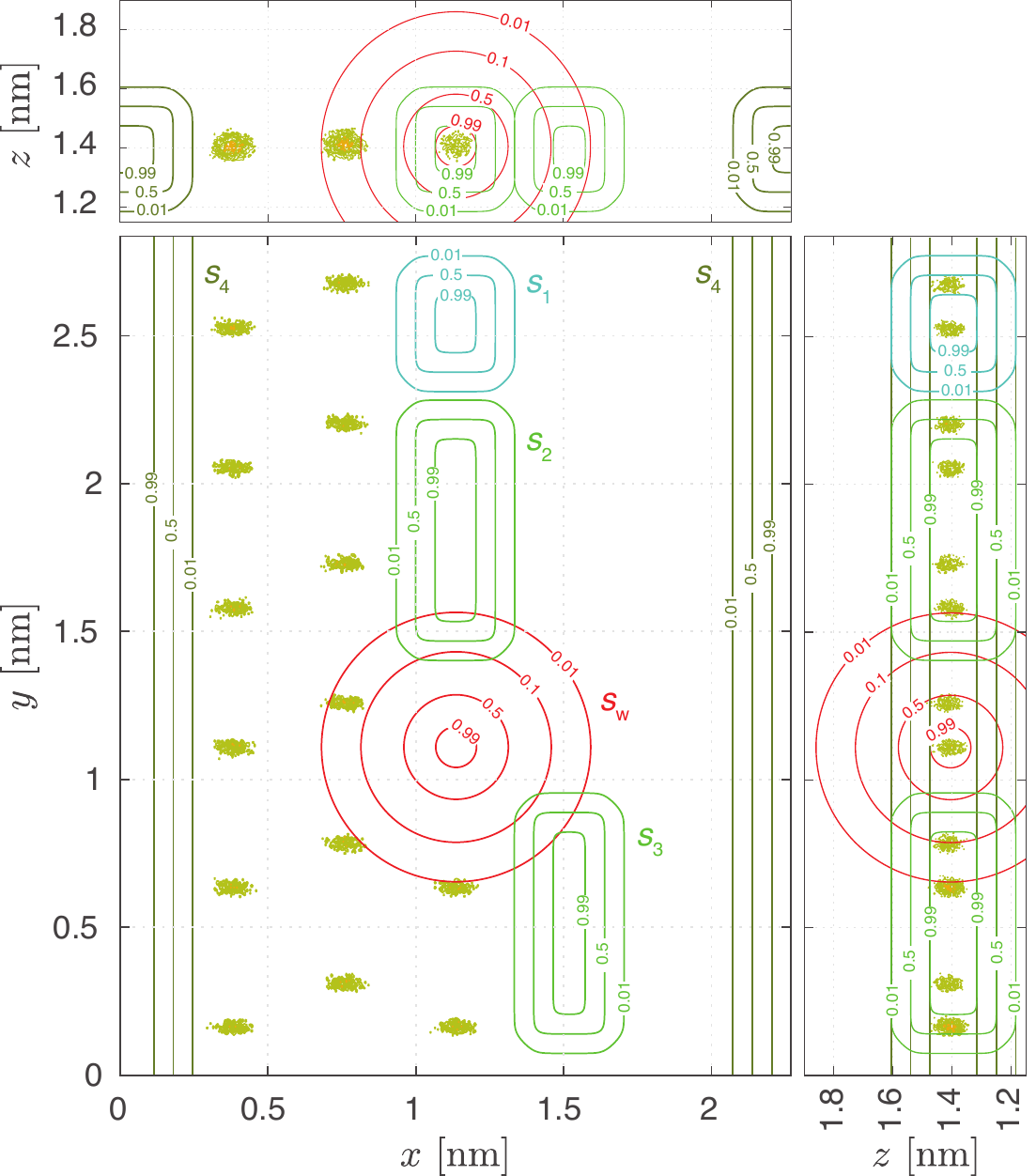}
\end{center}
\caption{Contour lines of adsorption site CVs and histogram of urea carbon atom positions of the unfinished surface layer.} \label{fig:walls}
\end{figure}

The crystal surface with the unfinished layer is comprised of other sites than only the biased kink site: the kink site opposite to the biased kink site and edges (see Figure \ref{fig:walls}). For the relatively fast growing crystal systems urea and naphthalene, it can happen that the edges and kink sites are growing on other sites and not only on the biased kink site.

Similarly as we did with the surface structure CV to prevent molecules from dissolving, wall potentials are introduced along the edges through adsorption site CVs to prevent the growth of these sites. In this work these CVs, $s_\kappa$, are comprised of simple logistic switching functions in all three spatial directions. For solute molecule $i$ $s_\kappa$ is defined as:
\begin{align}
s_\kappa = \sum_k \bigg[ \frac{1}{1 + \exp (-\sigma_\kappa (x_k - x_{\text{l}\kappa}))} &\left( 1 - \frac{1}{1 + \exp (-\sigma_\kappa (x_k - x_{\text{u}\kappa}))} \right) \\
\cdot \frac{1}{1 + \exp (-\sigma_\kappa (y_k - y_{\text{l}\kappa}))} &\left( 1 - \frac{1}{1 + \exp (-\sigma_\kappa (y_k - y_{\text{u}\kappa}))} \right) \\
\cdot \frac{1}{1 + \exp (-\sigma_\kappa (z_k - z_{\text{l}\kappa}))} &\left( 1 - \frac{1}{1 + \exp (-\sigma_\kappa (z_k - z_{\text{u}\kappa}))} \right) \bigg],
\end{align}
where $\sigma_\kappa$ is the steepness of the logistic functions. $x_{\text{l},\kappa}$, $y_{\text{l},\kappa}$, $z_{\text{l},\kappa}$ define the lower bounds and $x_{\text{u},\kappa}$, $y_{\text{u},\kappa}$, $z_{\text{u},\kappa}$ the upper bounds of the intervals (in each spatial direction) in which the adsorption site CV should act on the molecule center position $x_i$, $y_i$, $z_i$. $s_{\kappa}$ is obtained by summing $s_{\kappa,i}$ over all solute molecules (excluding bulk- and unfinished surface layer molecules).

The harmonic wall potential is defined as:
\begin{equation}
V_\kappa =
\begin{cases}
0, & \text{if}\ s_\kappa < s_{\kappa,0}, \\
k_\kappa (s_\kappa - s_{\kappa,0})^2, & \text{else},
\end{cases}
\end{equation}
with force constant $k_\kappa$ and threshold $s_\kappa$ above which $V_\kappa$ is active.

Figure \ref{fig:walls} shows the contour lines of the adsorption site CVs (violet and green lines). The parameters used for the different systems are summarized in Table \ref{tab:walls}.

\begin{table}
\caption{Values of adsorption site CVs. The coordinates origin is set to the simulation box center.}
\centering
\begin{tabular}{ccc ccc ccc cc}
\toprule
&                    & & \multicolumn{3}{c}{urea} & & \multicolumn{2}{c}{naphthalene} \\
&                    & & MeCN   & EtOH   & MeOH   & & EtOH  & MePh      \\
\midrule 
\multirow{9}{*}{$s_1$}
& $\sigma_1$ [-]   & & 80     & 80     & 80     & & 80    & 80          \\
& $x_\text{l1}$ [nm] & & 0.8356 & 2.0742 & 2.0742 & & 0.06  & 1.2931    \\
& $x_\text{u1}$ [nm] & & 1.1356 & 0.1558 & 0.1558 & & 0.28  & 1.5131    \\
& $y_\text{l1}$ [nm] & & 1.9727 & 0.9620 & 0.9620 & & 1.10  & 1.1205    \\
& $y_\text{u1}$ [nm] & & 2.4789 & 1.3420 & 1.3420 & & 1.45  & 1.4924    \\
& $z_\text{l1}$ [nm] & & 0.9405 & 1.0847 & 1.0847 & & 1.31  & $-$0.2679 \\
& $z_\text{u1}$ [nm] & & 1.2705 & 1.4147 & 1.4147 & & 1.64  & 0.2521    \\
& $k_1$ [kJ/mol]     & & 30     & 30     & 30     & & 30    & 30        \\
& $s_{1,0}$ [nm]     & & 0.2    & 0.2    & 0.2    & & 0.2   & 0.2       \\ \midrule
\multirow{9}{*}{$s_2$}
& $\sigma_2$ [-]     & & 80   & 80     & 80     & & 80    & 80        \\
& $x_\text{l2}$ [nm] & & 0.8356 & 2.1642 & 2.1642 & & 0.06  & 1.2931    \\
& $x_\text{u2}$ [nm] & & 1.1356 & 0.0958 & 0.0958 & & 0.28  & 1.5131    \\
& $y_\text{l2}$ [nm] & & 2.4789 & 0.0520 & 0.0520 & & 0.35  & 0.3176    \\
& $y_\text{u2}$ [nm] & & 3.9973 & 0.8620 & 0.8620 & & 1.06  & 1.0592    \\
& $z_\text{l2}$ [nm] & & 0.9405 & 1.0847 & 1.0847 & & 1.31  & $-$0.2679 \\
& $z_\text{u2}$ [nm] & & 1.2705 & 1.4147 & 1.4147 & & 1.64  & 0.2521    \\
& $k_2$ [kJ/mol]     & & 5      & 5      & 5      & & 5     & 5         \\
& $s_{2,0}$ [nm]     & & 1      & 1      & 1      & & 1     & 1         \\ \midrule
\multirow{9}{*}{$s_3$}
& $\sigma_3$ [-]   & & 80     & 80     & 80     & & 80    & 80        \\
& $x_\text{l3}$ [nm] & & 0.4582 & 0.2801 & 0.2801 & & 0.45  & 1.6804    \\
& $x_\text{u3}$ [nm] & & 0.6782 & 0.4801 & 0.4801 & & 0.71  & 1.9404    \\
& $y_\text{l3}$ [nm] & & 0.2489 & 1.4980 & 1.4980 & & 1.80  & 1.8020    \\
& $y_\text{u3}$ [nm] & & 2.4789 & 2.2980 & 2.2980 & & 2.65  & 2.6415    \\
& $z_\text{l3}$ [nm] & & 0.9405 & 1.0847 & 1.0847 & & 1.31  & $-$0.2679 \\
& $z_\text{u3}$ [nm] & & 1.2705 & 1.4147 & 1.4147 & & 1.64  & 0.2521    \\
& $k_3$ [kJ/mol]     & & 5      & 5      & 5      & & 5     & 5         \\
& $s_{3,0}$ [nm]     & & 1      & 1      & 1      & & 1     & 1         \\ \midrule
\multirow{9}{*}{$s_4$}
& $\sigma_4$ [-]     & & 80      & 80        & 80        & & 80        & 80        \\
& $x_\text{l4}$ [nm] & & 2.5552    & 0.6374    & 0.6374    & & 1.2092    & 2.4833    \\
& $x_\text{u4}$ [nm] & & 2.9552    & 0.8774    & 0.8774    & & 1.6200    & 2.9933    \\
& $y_\text{l4}$ [nm] & & $-\infty$ & $-\infty$ & $-\infty$ & & $-\infty$ & $-\infty$ \\
& $y_\text{u4}$ [nm] & & $\infty$  & $\infty$  & $\infty$  & & $\infty$  & $\infty$  \\
& $z_\text{l4}$ [nm] & & 0.9405    & 1.0847    & 1.0847    & & 1.120     & $-$0.2679 \\
& $z_\text{u4}$ [nm] & & 1.2705    & 1.4147    & 1.4147    & & 1.760     & 0.2521    \\
& $k_4$ [kJ/mol]     & & 30        & 30        & 30        & & 30        & 30        \\
& $s_{4,0}$ [nm]     & & 1         & 1         & 1         & & 1         & 1         \\ \midrule
\multirow{9}{*}{$s_5$}
& $\sigma_4$ [-]   & & 80        & 80        & 80        & & -         & -         \\
& $x_\text{l4}$ [nm] & & $-\infty$ & $-\infty$ & $-\infty$ & & -         & -         \\
& $x_\text{u4}$ [nm] & & $\infty$  & $\infty$  & $\infty$  & & -         & -         \\
& $y_\text{l4}$ [nm] & & $-\infty$ & $-\infty$ & $-\infty$ & & -         & -         \\
& $y_\text{u4}$ [nm] & & $\infty$  & $\infty$  & $\infty$  & & -         & -         \\
& $z_\text{l4}$ [nm] & & $-$1.35   & 1.0847    & 1.0847    & & -         & -         \\
& $z_\text{u4}$ [nm] & & $-$0.90   & 1.4147    & 1.4147    & & -         & -         \\
& $k_5$ [kJ/mol]     & & 40        & 40        & 40        & & -         & -         \\
& $s_{5,0}$ [nm]     & & 0         & 0         & 0         & & -         & -         \\
\bottomrule
\end{tabular}
\label{tab:walls}
\end{table}

For the reweighting on the crystallinity CVs, only $V_2$ and $V_3$ (shown in green in Figure \ref{fig:walls}) were considered. $V_1$ and $V_4$ were not included because they are distant enough from the kink site and are addressing sites caused by the PBC and are not relevant for the kink growth process. Reweighting on $V_2$ and $V_3$ changes $\Delta F$ less than 0.5 kJ/mol (which is in the order of the overall accuracy of the WTMetaD sampling). The use of $V_{2}$ and $V_{3}$ is not necessary for simulations performed at undersaturated or around saturated conditions.
Most APIs will not need these walls because their growth is kinetically hindered such that kink growth events are rare within the simulation time span of $\sim$ 1 $\mu$s.

\subsection*{Crystallinity CVs}

\begin{figure}[!htbp]
\begin{center}
\includegraphics[width=7cm]{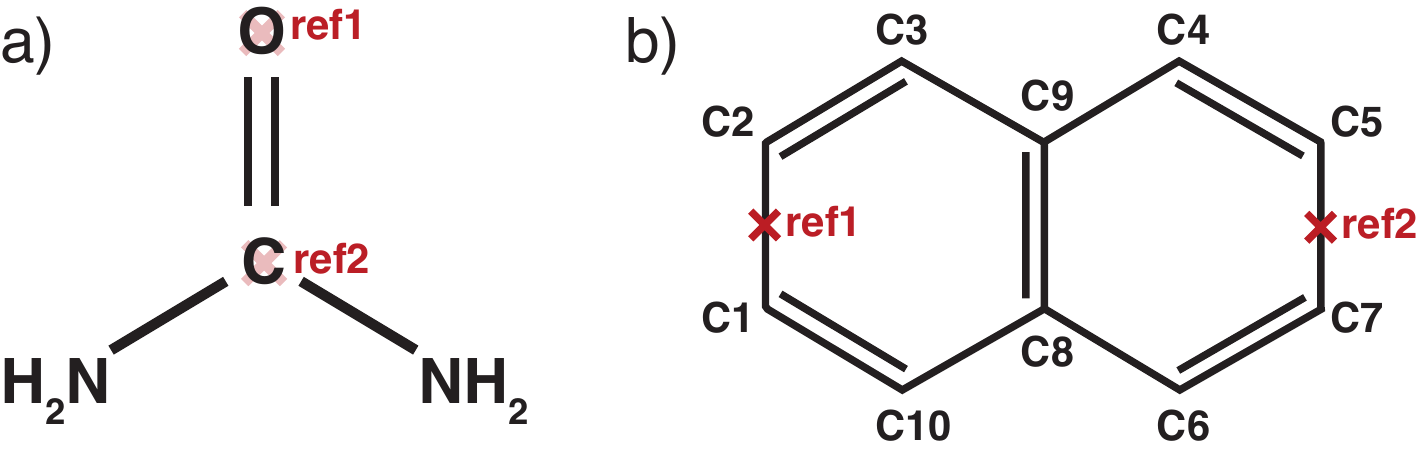}
\end{center}
\caption{Reference positions of the crystallinity CVs, $s_\text{c1}$ and $s_\text{c2}$. a) Urea: oxygen and carbon atom positions. b) Naphthalene: center of mass of carbon atom pairs C1-C2 and C5-C7 (which are interchangeable due to symmetry).} \label{fig:reference}
\end{figure}

\begin{figure}[!htbp]
\begin{center}
\includegraphics[width=\textwidth]{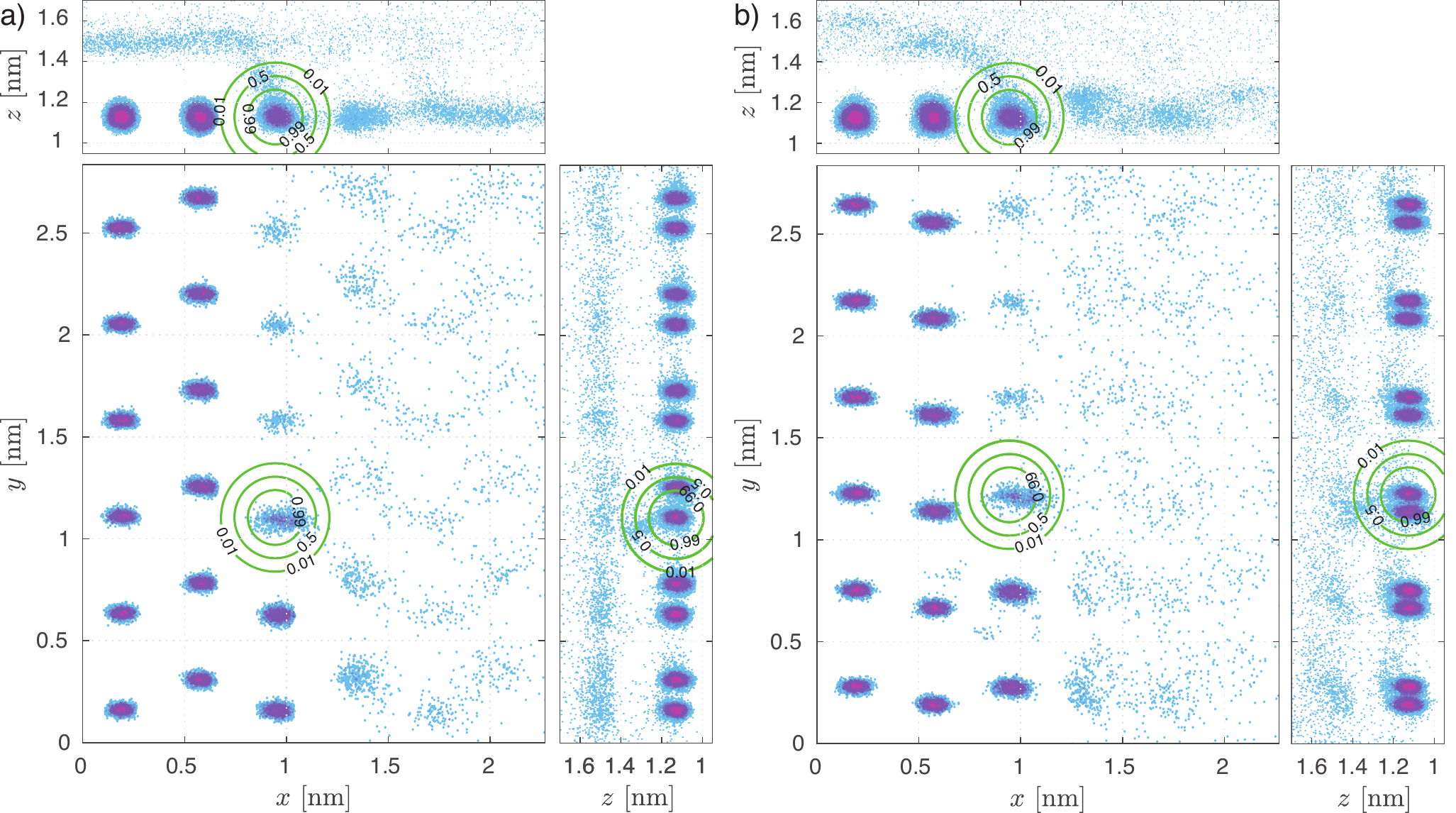}
\end{center}
\caption{Histogram of urea atom positions of the crystal surface layer together with the contour lines of the crystallinity CVs (green lines) projected along all three spatial directions. a) Carbon atoms and $s_\text{c1}$ contours. b) Oxygen atoms and $s_\text{c2}$ contours.} \label{fig:reweight}
\end{figure}

The reference atoms used for the crystallinity CVs, $s_\text{c1}$ and $s_\text{c2}$, of urea and naphthalene are shown in Figure \ref{fig:reference}.
The parameter values of $s_\text{c1}$ and $s_\text{c2}$ are listed in Table \ref{tab:reweight} and are chosen such that their values close to 1 correspond to a fully crystalline molecule at the biased kink site and values around 0 correspond to a fully dissolved biased kink site. Figure \ref{fig:reweight} shows the contour lines of the crystallinity CVs for the case of urea. The graph shows the histogram of the carbon atom positions (left) and the oxygen atom positions (right) at the crystal surface together with the contour lines of $s_\text{c1}$ and $s_\text{c2}$, which take the crystalline carbon atom position and oxygen atom position respectively at the biased kink site as references. The values of $s_\text{c1}$ and $s_\text{c2}$ which are between 0 and 1 correspond roughly to the urea atom positions within the region of transition, which exhibit the lowest density in the histogram in the surroundings of the biased kink site. The region of transition coincides approximately with the space between the nearest neighbors.

It suffices to take only the solute positions at the biased kink site into consideration for the crystallinity CVs, while neglecting the solvent, since the states of the biased kink site containing vacuum are very short lived and immediately refilled either with solvent or solute.

\begin{table}
\caption{Values of crystallinity CVs used in the reweighting.}
\centering
\begin{tabular}{cccccccccccc}
\toprule
&                             & & \multicolumn{3}{c}{urea} & & \multicolumn{2}{c}{naphthalene} \\
&                             & & MeCN  & EtOH    & MeOH   & & EtOH   & MePh    \\
\midrule
\multirow{5}{*}{$s_\text{c1}$}
& $r_\text{c1}^{(x)}$ [nm]    & & 2.8366 & 1.1369 & 1.1384 & & 1.8996 &  3.1273 \\
& $r_\text{c1}^{(y)}$ [nm]    & & 2.2038 & 1.1095 & 1.1091 & & 1.3968 &  1.3919 \\
& $r_\text{c1}^{(z)}$ [nm]    & & 1.1404 & 1.2844 & 1.4044 & & 1.2767 & $-$0.0964 \\
& $\sigma_\text{c1}$ [-]      & & 70     & 70     & 70     & & 70     & 70      \\
& $d_\text{c1}$ [nm]          & & 0.15   & 0.15   & 0.15   & & 0.20   & 0.20    \\ \midrule
\multirow{5}{*}{$s_\text{c2}$}
& $r_\text{c2}^{(x)}$ [nm]    & & 2.8366 & 1.1372 & 1.1383 & & 1.7266 &  2.9597 \\
& $r_\text{c2}^{(y)}$ [nm]    & & 2.2038 & 1.2304 & 1.2296 & & 1.3119 &  1.3049 \\
& $r_\text{c2}^{(z)}$ [nm]    & & 1.1404 & 1.2827 & 1.4027 & & 1.7121 &  0.3338 \\
& $\sigma_\text{c2}$ [-]      & & 70     & 70     & 70     & & 70     &  70     \\
& $d_\text{c2}$ [nm]          & & 0.15   & 0.15   & 0.15   & & 0.20   &  0.20   \\
\bottomrule
\end{tabular}
\label{tab:reweight}
\end{table}

\section*{Sampling of solubility with chemically distinct kink sites}

\begin{figure}[!htbp]
\begin{center}
\includegraphics[width=\textwidth]{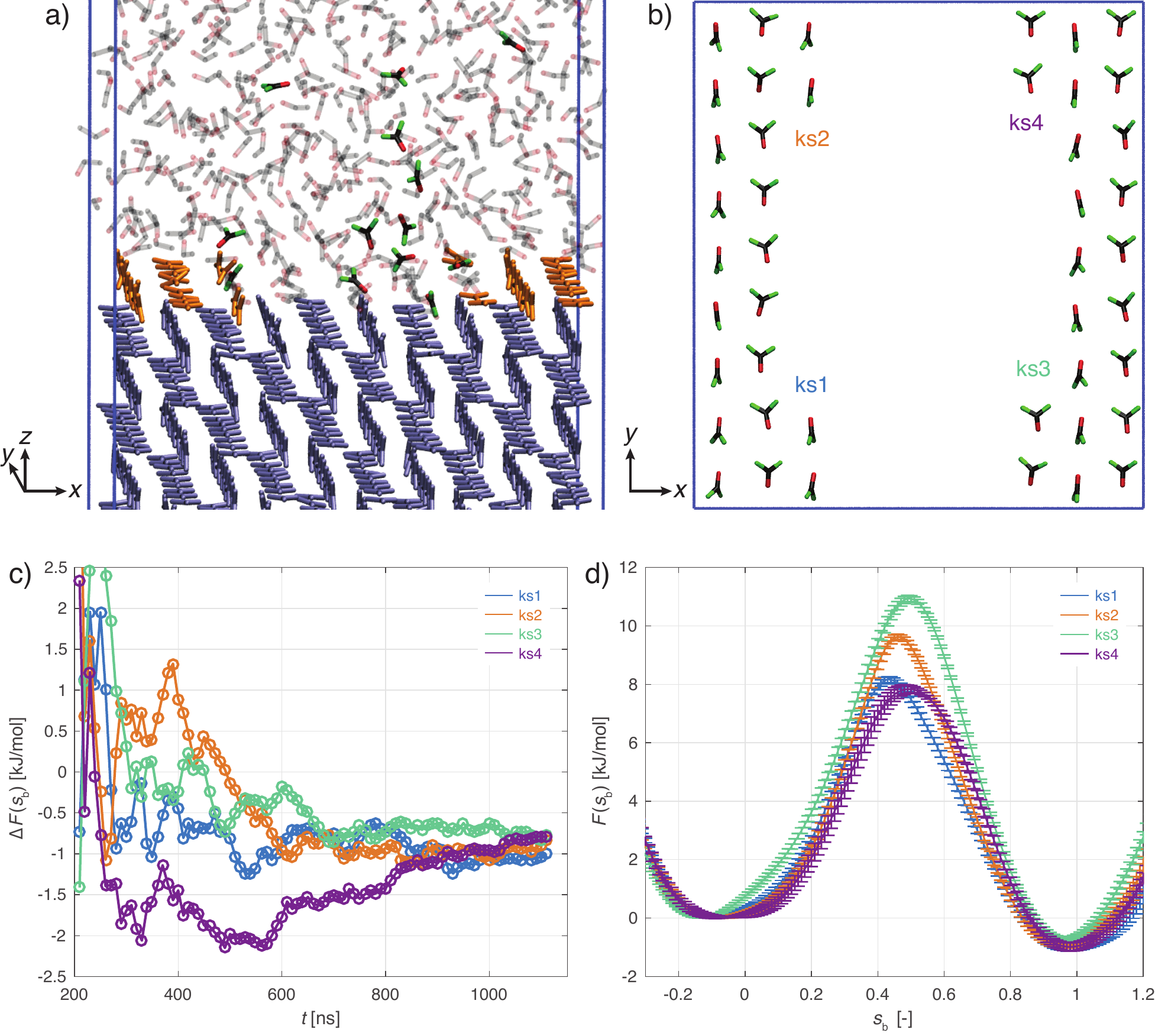}
\end{center}
\caption{Biased simulations of urea grown in ethanol for four chemically distinct kink sites: a) Visualization of the crystal surface of the simulation box. b) Unfinished surface layer with labeled kink sites ks1-4. c) Time evolution of the energy difference of the grown and dissolved states $\Delta F (s_\text{b1-4})$ reweighted over the biased CVs, $s_\text{b1-4}$. d) Corresponding $F(s_\text{b1-4})$ averaged over the last 200 ns.} \label{fig:aug_setup}
\end{figure}

$\Delta F$ should not depend on the kink site, as long as the growth unit corresponds to a single molecule along the edge of interest. To quantitatively verify this assumption we have run kink growth simulations of the four chemically distinct kink sites of urea face $\{110\}$ with the unfinished surface layer cut along edge $[001]$ and grown in ethanol at a solute mole fraction of $x = 0.020$. All four kink sites were biased simultaneously.
The visualization of the crystal surface is shown in Figure \ref{fig:aug_setup}a) and the unfinished surface layer with labeled kink sites ks1-4 is shown along the $z$-axis in Figure \ref{fig:aug_setup}b).

The simulation convergence of growth and dissolution of the kink sites is presented in Figure \ref{fig:aug_setup}c), which shows the time evolution of the energy difference of the grown and dissolved states $\Delta F (s_\text{b1-4})$ reweighted over the biased CVs, $s_\text{b1-4}$. The corresponding $F(s_\text{b1-4})$ averaged over the last 200 ns are shown in Figure \ref{fig:aug_setup}d), which clearly show that the solubility is the same for all 4 kink sites (the energy differences are within the accuracy of the method of $\sim$ 0.5 kJ/mol). The values of $\Delta F$, obtained with reweighting on the crystallinity CVs, are: $\Delta F_\text{ks1} =$ 0.05 kJ/mol, $\Delta F_\text{ks1} =$ -0.03 kJ/mol, $\Delta F_\text{ks1} =$ 0.04 kJ/mol, and $\Delta F_\text{ks1} =$ 0.33 kJ/mol.
These values are in agreement with the ones reported in the main manuscript, which were obtained with a smaller simulation box setup.

It is interesting to note, that the activation energy barriers of $F(s_\text{b1-4})$ in Figure \ref{fig:aug_setup}d) are smaller for kink sites, which face an oxygen atom of the unfinished row (ks1 and ks4) while the activation energies of the kink sites facing amine groups of the unfinished row are slightly larger (ks2 and ks3).

\end{document}